\documentclass[draft]{agujournal}
\usepackage{natbib}
\usepackage{amsmath}
\usepackage{graphicx}
\usepackage{siunitx}
\usepackage{booktabs}
\usepackage{todonotes}
\usepackage{setspace}

\graphicspath{{figs/}}
\newcommand{\qt}{$q_T$ }
\newcommand{\sli}{$s_L$ }
\newcommand{\norm}[1]{\left\lVert#1\right\rVert}

\makeatletter
\renewcommand{\todo}[2][]{\@todo[caption={#2}, #1]{\begin{spacing}{0.5}#2\end{spacing}}} 
\makeatother 

\journalname{Journal of Advances in Modeling Earth Systems (JAMES)}
\begin{document}
%
%


\title{Spatially Extended Tests of a Neural Network Parametrization Trained by
  Coarse-graining}

%
%




\authors{Noah D. Brenowitz\affil{1}\thanks{Department of Atmospheric Sciences, University of Washington, Box 351640, Seattle, WA 98195-1640}, Christopher S. Bretherton\affil{1}}

\affiliation{1}{Department of Atmospheric Sciences, University of Washington}





\correspondingauthor{Noah Brenowitz}{nbren12@uw.edu}




\begin{keypoints}
\item  A neural network parametrization is coupled to a GCM
\item The neural network is trained by coarse-graining a near-global cloud-system resolving simulation 
\item The coupled simulation accurately forecasts the weather of the training data 
\end{keypoints}

%
%


\begin{abstract}
General circulation models (GCMs) typically have a grid size of 25--200~km.  Parametrizations are used to represent diabatic processes such as radiative transfer and cloud microphysics and account for sub-grid-scale motions and variability.
Unlike traditional approaches, neural networks (NNs) can readily exploit recent observational
  datasets and global cloud-system resolving model (CRM) simulations to learn subgrid variability.
  This article describes an NN parametrization trained by coarse-graining a near-global CRM simulation with a 4~km horizontal grid spacing. 
  The NN predicts the residual heating and moistening averaged over
  $(\SI{160}{km})^2$ grid boxes as a function of the coarse-resolution fields within
  the same atmospheric column.
  This NN is coupled to the dynamical core of a GCM with the same \SI{160}{km} resolution.
  A recent study described how to train such an NN  to be numerically stable  when coupled to specified time-evolving advective forcings in a single column model,
  but feedbacks between NN and GCM components cause spatially-extended
  simulations to crash within a few days.
  Analyzing the linearized response of such an NN reveals that it learns to exploit a
  strong synchrony between precipitation and the atmospheric state above
  \SI{10}{km}.
  Removing these variables from the NN's inputs stabilizes the coupled simulations, which predict the future state more accurately than a coarse-resolution simulation without any parametrizations of sub-grid-scale variability, although the mean state slowly drifts.
\end{abstract}


%
%

\section{Introduction}

Current global climate and weather models cannot explicitly resolve many
important physical processes because of computational expenses.
Global weather and short-range climate forecast models support grid sizes of \SIrange{10}{50}{km} \citep{ECMWF2018,NOAA2018}, while current climate models typically have \SIrange{0.25}{1}{\deg} grids.
Unresolved processes, include cumulus convection, turbulence, and subgrid cloud variability are approximated by sub-grid-scale parametrizations \citep{Palmer2001-rg}. 
These are usually designed by human physical intuition, informed by process modeling and observations.

Cumulus convection in the tropics is one of the most dynamically important
processes in the atmosphere yet it is extremely difficult to parameterize because
of the multiscale nature of moist flows \citep{Nakazawa1988-xh, Majda2007-qr}.
There is a long-standing debate about whether convective parametrizations should be based on moisture convergence or quasi-equilibrium closures \cite{Arakawa2004-io}.
Moreover, important mean-state biases in climate modeling such as the double-ITCZ
bias are sensitive to convective parametrization \citep{Zhang2006-iy,Woelfle2018-gi}.
Climate models also struggle to simulate observed aspects of tropical variability such as the
diurnal cycle of continental precipitation \citep{Stratton2012-iu} and the Madden Julian Oscillation (MJO) \citep{Jiang2015-jo, Jiang2017-sr}.  Thus, an improved convective parametrization could help improve weather and climate simulations.

With sufficient computational resources, cloud-system resolving models (GCRMs) with sub-\SI{5}{km} grid spacing can be run in global domains without deep convective parametrizations  \citep{Satoh2008-pt,Bretherton2015-iz}.
Recent studies have used such simulations to study the moist static energy
budget \citep{Bretherton2015-iz} and cloud feedbacks \citep{Narenpitak2017-ep}.
The DYAMOND project recently compared 40-day hindcasts by 10 independently-developed global CRMs (CRMs), showing reassuring similarity in their simulated patterns of precipitation and high cloud \citep{Stevens2019}.
Centennial-scale climate GCRM simulations are not yet feasible, so improving parametrizations in coarse-resolution models remains an important goal.  

It remains challenging for human experts to translate the vast volume of information in new high-resolution datasets and models into better parametrizations of moist physical processes incorporating subgrid variability.
On the other hand, parametrizations can also be built using powerful regression
tools developed in the field of machine learning (ML) \citep{Rasp2018-ff,
  OGorman2018-hn, Brenowitz2018-td}.
In particular, \citet{Brenowitz2018-td} trained a neural network (NN) \citep{Goodfellow2016-wf} parametrization for the combined diabatic heating and moistening in a coarse-grid model by
coarse-graining a GCRM simulation.  They showed that coupling this parametrization to specified dynamical tendencies in a single column model gives stable and highly accurate simulations if the NN is trained to optimize model performance over multiple time steps.

We now extend our single column model results by coupling an NN parametrization to the dynamical core of a GCM.
As in \citet{Brenowitz2018-td}, this unified physics scheme will predict the sources of water and heat due to latent heating, turbulence, radiation, and any other physical process beyond resolved advection.
Our main goal is to produce accurate multiple day forecasts with the coupled GCM-NN model. 

After a brief survey of recent work on ML parametrization in Section
\ref{sec:lit-review}, we describe our training data and GCM configuration in Section \ref{sec:gcm}. 
Section \ref{sec:methods} describes our coarse-graining and machine learning strategies. 
Particular focus is placed on the difficult task of ensuring numerical stability in spatially
extended simulations (Section \ref{sec:coupled-instability}).
Section \ref{sec:results} shows the results of the coupled simulations, and
we conclude in Section \ref{sec:conclusion}.

\section{Review of Machine Learning Parametrization}
\label{sec:lit-review}


One attractive use of machine learning (ML) is to automatically tune existing GCM parametrizations, which build in physical insights and constraints that may help them apply across a range of climates.
Proposed techniques include data assimilation 
\citep{Schneider2017-qq,Lyu_undated-qw} and genetic algorithms \citep{Langenbrunner2017-ed}.
These techniques can tune a few free parameters, but may not scale to larger numbers of parameters.
Moreover, existing parametrizations may not be flexible enough to be realistic
in part because they have so few parameters.
For instance, adjusting the entrainment rates in a convection scheme can improve
mean-state bias, but harm the variability \citep{Kim2011-de,Mapes2011-vh}.

A more ambitious approach replaces an existing GCM parametrization
with a machine learning (ML) model, which has enough parameters to capture arbitrarily complex relationships \citep{Cybenko1989-hr,Hanin2017-hu} present within a training dataset.
ML models are typically trained by minimizing a loss function, such as the mean-squared error (MSE) compared to some reference outputs from the training data; the choice of the loss function is subjective and a key to good performance.
Early studies \citep{Chevallier1998-su,Krasnopolsky2005-ca} trained NNs \citep{Goodfellow2016-wf} to emulate the outputs of radiative transfer codes in order to decrease computational expense.
Subsequent studies also trained NNs to emulate existing convective
parametrizations \citep{OGorman2018-hn} and super-parametrizations (SP)
\citep{Rasp2018-ff}.
In each of these instances, the ML model is trained with a nearly ideal dataset,
where the inputs are the state of the atmosphere, and the output is a tendency
that has actually driven a GCM.

Training a parametrization by coarse-graining a GCRM is a more difficult problem because it lacks this hierarchical
structure.
On the other hand, GCRMs are the most realistic models available because they do not impose an arbitrary scale separation as SP does.
In this setting, how does one define the target tendency?
In the first study of its kind, \citet{Krasnopolsky2010-nn, Krasnopolsky2013-zw} defined the output as the residual heating and moistening \citep{Yanai1973-gt} in limited area simulations.
They diagnosed heating rates and cloud fractions with high accuracy but did not
present any prognostic simulations. 
\citet{Brenowitz2018-td} extended this work in two ways. 
First, their training dataset was a near-global CRM with rich multiscale convective organization.
Second, they found single column model (SCM) simulations with NNs trained in this way diverged to infinity after just a few time steps.  
They ensured long-term stability by minimizing the loss accumulated over several predicted time steps.

In summary, in coarse-graining, there is no clear target to emulate. While \citet{Brenowitz2018-td} developed an approach for ensuring numerical stability in a single-column mode with prescribed dynamics, they did not test it within a full three-dimensional GCM where the dynamics interact with the NN parametrization.

\section{Training Data and Atmospheric Model Configuration}
\label{sec:gcm}

\subsection{Training Data}

We use the same training dataset as \citet{Brenowitz2018-td}: a
near-global aquaplanet simulation (NG-Aqua) using the System for Atmospheric
Modeling (SAM) version 6.10 \citep{Khairoutdinov2003-du}. 
SAM is run in a cloud-system resolving configuration with a horizontal grid spacing of
\SI{4}{km} in a tropical channel domain measuring \SI{20480}{km} zonally by
\SI{10240}{km} meridionally. 
The simulation has 34 vertical levels with a spacing that increases from
\SI{75}{m} at the surface to \SI{1200}{m} in the troposphere and stratosphere.
The atmospheric circulation is driven by a zonally symmetric sea surface
temperature (SST) profile which peaks at \SI{300.15}{K} at the equator
($y=\SI{5120}{km}$) and decreases to \SI{278.15}{K} at the poleward boundaries. 
Because SAM was originally designed for small-scale modeling, the grid is
Cartesian and no spherical effects are included, except for a meridionally
varying Coriolis parameter.
Despite the simplifications described above, NG-Aqua features realistic multiscale organization of tropical and midlatitude systems.

The simulation uses the radiation scheme from version 3 of the Community
Atmosphere Model (CAM) with a zonally symmetric diurnal cycle and a bulk
microphysics scheme \citep{Khairoutdinov2003-du}.
It was initialized with small random perturbations to the temperature, spun up on a 20~km grid, then interpolated to 4~km and 
run for 100 days, storing the full 3-D outputs every 3 hours.
To allow for the circulation to statistically equilibrate at \SI{4}{km}, all analyses described below are
performed on the final 80 days.
More details about the model configuration are described by \citet{Brenowitz2018-td} and \citet{Narenpitak2017-ep}.

\subsection{Coarse-resolution model configuration}

We test the NN schemes within an atmospheric model with a grid resolution
of \SI{160}{km}, which is within the range of modern GCMs.   This scale is coarse enough that the grid-mean precipitation has significant auto-correlation over the 3-hour sampling interval of the
training data, which should therefore sufficiently resolve the time-evolving grid-mean moist convective dynamics.

We use SAM rather than a more traditional GCM as our coarse-grid model to ensure that
it has the same geometry and dry dynamics as the training data.
This model, coarse-SAM (cSAM), is run with the same vertical grid as the NG-Aqua simulation. Its default microphysics scheme has the same three prognostic thermodynamic variables as SAM:
the total non-precipitating water mixing ratio $q_T$, the precipitating water mixing ratio $q_p$, which is a fast
variable that we will neglect in the parametrizations below; and the liquid-ice static energy $s_L$.
See \citep[Appendix A]{Khairoutdinov2003-du} for more details.

SAM is typically used for cloud-resolving simulations so
running it efficiently and stably with coarse resolution required several
modifications.
Most importantly, horizontal hyper-diffusion $- K_4 \nabla^4$ of $s_L$, $q_T$, and the three wind components, $u$, $v$, and $w$, was added to suppress grid-scale
oscillations and model blow-up (divergence of the solutions to machine infinity in a finite amount of time).
For the 160~km grid, $K_4= \SI{1e16}{m^4/ s}$ was sufficient to damp grid-scale
oscillations.

For NN simulations, we use a simplified momentum damping rather than SAM's default turbulence closure, which severely limits the time step on the 160~km grid.
This forcing damps the velocity towards zero at a rate which decreases linearly
from $k_f=\SI{1}{\per\day}$ at the surface to zero at a $\sigma$-level of $0.7$ \citep{Held1994-py}. Because this is different than NG-Aqua, it leads to an inevitable drift of cSAM simulations away from NGAqua over the course of a few days.  In the future, we hope to use learned coarse-grained momentum tendencies from NG-Aqua in place of this approach.

We compare simulations with cSAM coupled to our NN parametrization with a base version of cSAM with the microphysics and radiation tendencies calculated from grid-scale mean thermodynamics. 
Unlike the NN simulations, the surface fluxes are computed interactively and coupled to SAM's default turbulence scheme.
The base cSAM will only precipitate when a coarse-resolution grid-cell achieves saturation, so we expect it will produce noisy simulations.
Ideally, we would have preferred to compare the NN parametrizations
against a traditional GCM parametrization suite, as did \citet{Rasp2018-ff} and
\citet{Brenowitz2018-td}, but no such scheme has been implemented in SAM.

In this article, we assess the accuracy of 10-day weather simulations.  They are
initialized with the coarse-grained outputs from a particular time, day 100.625, of the NG-Aqua simulation.  Our goal is to obtain cSAM simulations that best match the actual evolution of NG-Aqua over the following 10 days.
All of the simulations presented in this paper use a \SI{120}{s} timestep.
The thermodynamic variables and vertical velocity from the 4 km NG-Aqua data,
are averaged over $40^2$ grid cell blocks, but the horizontal velocities must be
averaged along the interfaces of these blocks to respect the anelastic mass
conservation equation.

\section{Machine Learning Parametrization}
\label{sec:methods}

\subsection{Coarse-graining Problem}

\newcommand{\x}{\mathbf{x}}
\newcommand{\y}{\mathbf{y}}

An NN will parameterize the unknown sources of the thermodynamic variables
$q_T$ and $s_L$. Their coarse-resolution budgets are given by
\begin{align}
  \frac{\partial \overline{s_L}}{\partial t} &= \left( \frac{\partial\overline{ s_L}}{\partial t} \right)_{\text{GCM}} + Q_1\label{eq:q1} \\ 
  \frac{\partial \overline{q_T}}{\partial t} &= \left( \frac{\partial\overline{ q_T}}{\partial t} \right)_{\text{GCM}} + Q_2 \label{eq:q2}
\end{align}
where $\overline{f}$ is the horizontal average of $f$ over the coarse-grid
boxes. 
The first term on the right-hand side of these budgets are the tendencies due to
the GCM, which includes advection and any other explicitly treated process. 
The apparent heating $Q_1$ and moistening $Q_2$ are defined as a budget residual
from these GCM tendencies \citep{Yanai1973-gt}. 
Because they are residuals, they contain the effects of all unresolved and
untreated physics including latent heating, turbulent mixing, and radiation.

The goal of any parametrization is to approximate $Q_1$ and $Q_2$ as potentially
stochastic and non-local functions of the coarse resolution variables alone \citep{Palmer2001-rg}.
Because radiative heating and moist physical processes couple the atmosphere more strongly in the vertical
than the horizontal direction, we follow the assumption of horizontal locality typically made in moist physics parametrizations; this reduces the dimensionality of the training problem, allowing robust results to be obtained from our NG-Aqua training dataset.  For simplicity, we also treat $Q_1$ and $Q_2$ as deterministic functions of the inputs.

The inputs are the vertical profiles of the prognostic variables $\overline{q_T}$ and $\overline{s_L}$, as well as the sea surface temperature (SST) and downwelling insolation (SOLIN) at the top of the atmosphere, which are external parameters.
Unlike \citet{Brenowitz2018-td}, we do not use latent heat flux because it depends on the surface winds, which tend to be inaccurate because they are forced by the simplified damping scheme described in Section \ref{sec:gcm} rather than a coarse-grained source from NG-Aqua.

Let the vector $\x^o_i$ concatenate the observed prognostic variables $\overline{q_T}$ and
$\overline{s_L}$ over all coarse-grained grid levels in grid column $i$, and let $\y^o_i$ contain the auxiliary variables SST and SOLIN at this location.
With these assumptions, \eqref{eq:q1} and \eqref{eq:q2} can be combined:
\begin{align}
   \frac{d \mathbf{x}^o_i}{dt} = g_i(\mathbf{x}^o, \y^o) + f(\mathbf{x}^o_i, \mathbf{y}^o_i; \theta)  +\epsilon_i, \label{eq:dxdt}
\end{align}
where the subscript $i$ is the index of the given horizontal grid cell and the superscript $o$ indicates that these are observed quantities.
The function $f$ and its parameters
$\theta$ are invariant for all locations.
Here, $g$ is the GCM tendency from coarse-grid advection (which involves other grid columns and hence is nonlocal); $f$ is the portion of the apparent sources $Q_1$ and $Q_2$ that can be modeled deterministically, which the NN will be used to parameterize; and $\epsilon_i$ includes stochastic and structural error. 
As with \citet{Brenowitz2018-td}, $f$ will be parameterized as an NN where the parameters $\theta$ include the weights and biases of the network's layers.

\subsection{Loss Functions for Numerically Stable Parametrizations}

How do we find the parameters $\theta$ from the NG-Aqua time series $\x^o(t)$?
\citet{Brenowitz2018-td} trained a numerically stable NN for use in single
column model (SCM) simulations by minimizing the multiple step prediction error (MSPE) over a 1-3 day prediction window $T$.
SCM dynamics decouple the dynamics of an atmospheric column from its surroundings and are described mathematically by 
\begin{equation}
  \label{eq:scm}
  \frac{d\x_i^{SCM}}{dt} = g_i(\x^o(t), \y^o(t)) + f(\x^{SCM}_i, \mathbf{y}^o_i(t); \theta),
\end{equation}
where $\x^{SCM}_i(t ;t_0, \theta)$ is the single column time series for a given
location $i$, starting prediction time $t_0$, and set of parameters $\theta$.
Compared to \eqref{eq:dxdt}, the global state $\x^o$ and the auxiliary variables $\y^o$ are prescribed functions of time, which are decoupled from the local dynamics.

\citet{Brenowitz2018-td} define the MSPE as 
\begin{equation}
  \label{eq:mspe}
  J_{\text{MSPE}}(\theta) = \sum_{i,t_0} \frac{1}{T} \int_{t_0}^{t_0+T} \norm{\x_i^{SCM}(t; t_0, \theta) - \x^o_i(t)}^2_M dt
\end{equation}
where the sum ranges over all possible prediction start times $t_0$ and spatial
locations $i$.
The inner norm is given by
\begin{equation}
  \label{eq:norm}
  \norm{
    \begin{bmatrix}
      \mathbf{s}\\
      \mathbf{q}
    \end{bmatrix}
}_M^2 =   \frac{\lambda_q}{M}\int q^2(z) \rho_0(z) dz + \frac{\lambda_s}{M}\int s^2(z) \rho_0(z) dz,
\end{equation}
where $\rho_0$ is SAM's reference density profile. 
$M=\int \rho_0 dz$ is the mass of an atmospheric column while $\lambda_s$ and
$\lambda_q$ are weight constants for the temperature and humidity, respectively. 
In this study, we find that when setting $\lambda_q= \SI{1.0}{kg^2/g^2}$ and
$\lambda_s = \SI{1.0}{K^{-2}}$ each variable contributes comparably to the overall loss.
This resulted in a highly accurate single column time-stepping scheme which
closely replicated the $q_T$ and $s_L$ time series of the NG-Aqua data.

\begin{figure}
  \includegraphics[width=\textwidth]{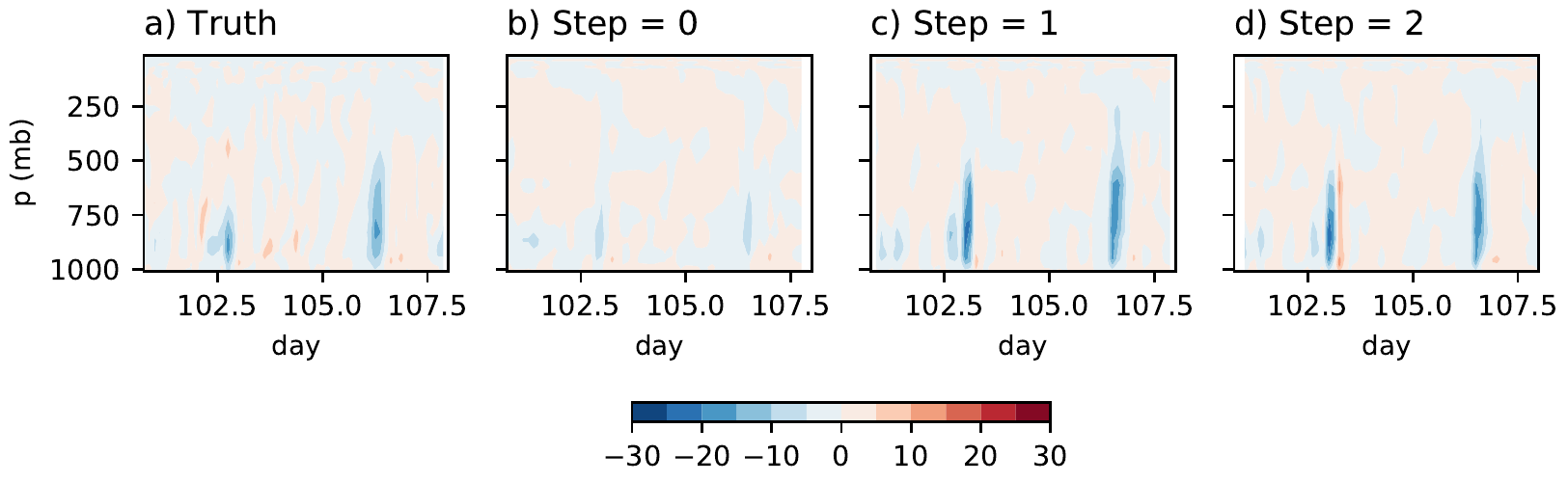}
  \caption{Time series of the $Q_2$ (g/kg/day) predicted by an NN trained to minimize
    $J_{MSPE}$. 
    These panels show the true value of $Q_2$ (a) and the predictions after a
    single column model has spun-up for the indicated number of 3 hour time
    steps (b-d).}
  \label{fig:spin-up-problem}
\end{figure}

Unfortunately, MSPE training produces NN schemes which implicitly depend
on the temporal discretization and time step used to simulate \eqref{eq:scm}.
The main symptom of this is a ``spin-up'' error where the SCM only produces
accurate tendencies after a few time steps of prediction (see Figure
\ref{fig:spin-up-problem}): the spun-up predictions (panel c, d) more closely resemble the truth (panel a) than the instantaneous prediction (b) does.
Because this spin-up process occurs after 3 hours, it did not cause large errors in the single column simulations of \citet{Brenowitz2018-td}.
However, for spatially-extended simulations, the coupling between the GCM dynamics and the NN occurs at every \SI{120}{s} time step, so the NN must produce accurate predictions from the start.
Thus, MSPE-trained NNs are not suitable for use within cSAM because they spin-up on a time-scale longer than the cSAM time step.

Our goal in this article is to develop a loss function which produces an NN that gives spatially-extended coarse-grid simulations that are stable, unbiased, and insensitive to the model time step.
This loss function is inspired by the Lax equivalence theorem in numerical analysis, which asserts that a finite difference method is accurate if and only if it is consistent and stable \citep{Quarteroni2007-cq}.
The loss is given by
\begin{equation}
  \label{eq:total}
  J_{\text{total}}(\theta) = J_{\text{instant}}(\theta) + J_{\text{stability}}(\theta).
\end{equation}
Consistency will mean that the $f$ approximates $Q_1$ and $Q_2$ instantaneously, and is accomplished by minimizing the loss given by
\begin{align}
  \label{eq:instant-loss} 
  J_{\text{instant}}(\theta) &= \sum_{i,n} \norm{f(\x^o_i(t^n), \y^o_i(t^n); \theta) -
    \begin{bmatrix}
      \mathbf Q_1\\
      \mathbf Q_2
    \end{bmatrix}
}_M^2, \\
  &= \sum_{i,n} \norm{\frac{d \x^o_i}{dt}(t^n) - \left[ g_i(\x^o(t^n),\y^o(t^n))  + f(\x_i^o(t^n), \y^o_i(t^n), \theta) \right]}^2_M.
\end{align}
In practice, the storage terms $\frac{d \x^o_i}{dt}$ are estimated from the 3 hourly sampled data using finite differences.
Thus, minimizing $J_{\text{instant}}$ demands that $f$ approximates the apparent heating and moistening.

\newcommand{\tavg}[1]{\langle #1 \rangle}

On the other hand, stability requires that the small perturbations to the inputs
$\mathbf{x}$ do not grow under the dynamics given by \eqref{eq:scm}.
In practice, we enforce this using a penalty given by
\begin{equation}
  J_{\text{stability}} =   \sum_{i,t_0} \norm{ \x^{mean}_i(t_0 + T; t_0, \theta) - \tavg{\x^o_i}}^2_M,
\end{equation}
where $\tavg{\cdot}$ is the time average operator. $\x^{mean}_i(t, t_0)$ is the single
column simulation under the action of the time-mean forcing, whose dynamics are given by
\begin{equation}
\frac{d\x^{mean}_i}{dt}= f(\x^{mean}_i, \y^o_i; \theta) + \tavg{g_i(\x^o(t))}. 
\end{equation}
Its initial condition is $\x^{mean}_i(t_0, t_0; \theta) = \x^o_i(t_0)$.
This penalty demands that the SCM simulation does not diverge too quickly for a prediction of length $T$ (\SI{2.5}{d}).
Because $J_{\text{stability}}$ uses constant forcing, it is less sensitive than the MSPE to the temporal discretization of \eqref{eq:scm}.
In practice, NNs trained to minimize $J_{total}$ avoid spin-up errors and accurately predict $Q_{1}$ and $Q_{2}$ at the initial time.
Like MSPE-trained NNs, they do produce stable, albeit less accurate, SCM simulations.
Since our main goal is to produce accurate coupled GCM-NN simulations this deteriorated SCM performance is acceptable.

\subsection{Coupled Numerical Instability}
\label{sec:coupled-instability}

\begin{figure}
  \includegraphics{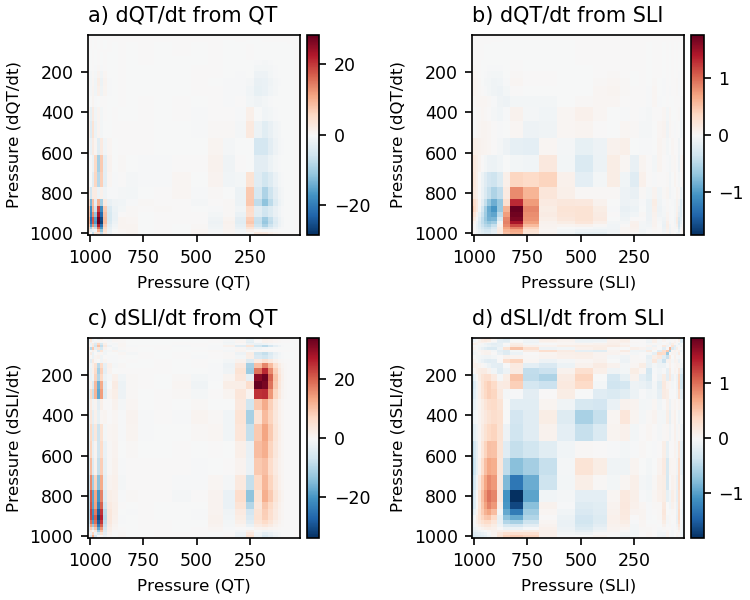}
  \caption{Linearized response of the neural network to input perturbations. 
    Compare to Figure 1 of \citet{Kuang2018-wh}.}
  \label{fig:saliency-bad}
\end{figure}

\begin{figure}
  \includegraphics[width=\textwidth]{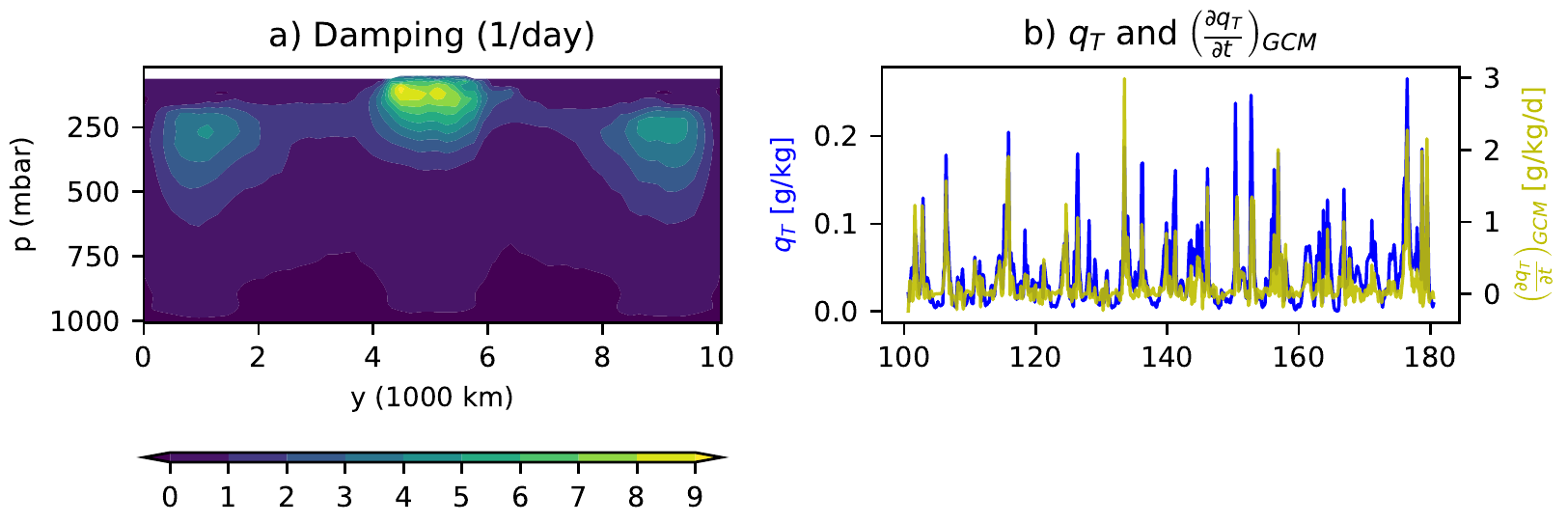}
  \caption{(a) Latitude-height structure of implied damping coefficient for the total water $q_T$ and (b) time
    series of $q_T$ with the corresponding dynamics tendency for an
    upper-level tropical location at $(x,y, p_0) = (\SI{0}{km}, \SI{5120}{km},
    \SI{192}{mb})$.
  }
  \label{fig:damping}
\end{figure}

\begin{figure}
  \includegraphics{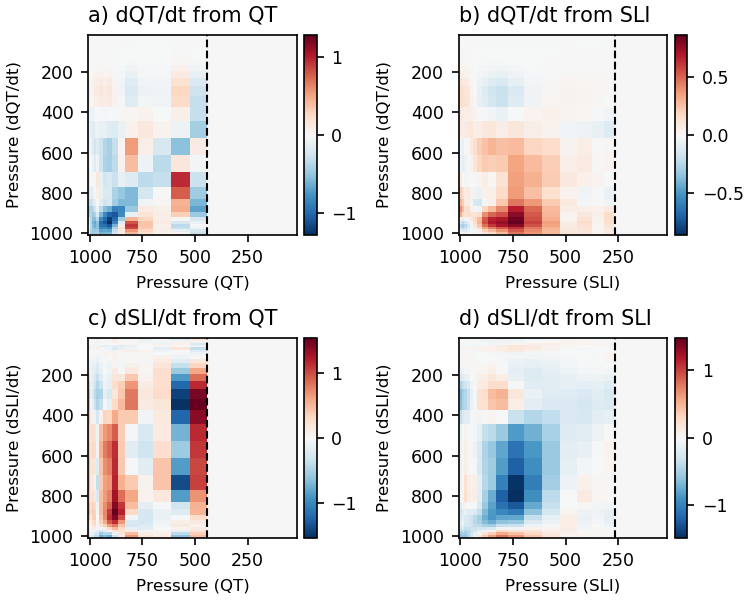}
  \caption{Same as Figure \ref{fig:saliency-bad}, but for a model which ignores
    the the humidity and temperature above the dashed lines at $z=\SI{450}{mb}$ and \SI{225}{mb},
    respectively. 
  }
  \label{fig:salience-good}
\end{figure}

The methods discussed still
do not prevent model blow-up when coupled to cSAM, as we will show in Section
\ref{sec:results}. 
This occurs because the single column training strategy does not account for feedbacks between cSAM and the NN so corresponding numerical instabilities are not penalized.

A similar feedback causes grid-scale storms when a GCM is coupled to a moisture-convergence closures \citep{Arakawa2004-io}.
\citet{Emanuel1994-fc} argue this closure fundamentally confuses cause and effect, and that precipitation drives moisture convergence rather than vice versa.
In fact, such misleading correlations are common in nonlinear dynamical systems such as the atmosphere, a phenomenon known as synchronization \citep{Pecora1997-kz}, so data-driven models must be taught to ignore such non-causal correlations.
Thus, we must determine if the NN is mistaking cause for effect.

\subsubsection{Interpreting the trained networks using linear response functions}

The linearized response of the NN to small perturbations can reveal if it is confusing cause and effect.
The linearized response function (LRF) is given by
\[J(\mathbf x, \y) = \nabla_{\mathbf x, \y}f(\mathbf x, \mathbf y; \theta),\]
where the derivative is with respect to the phase space variables $\mathbf x$.
For ease of discussion, we defer the model architecture and training details to Section \ref{sec:nn-details}.
We first train an NN using the stability-penalized loss in \eqref{eq:total} and then compute its LRF for the mean equatorial sounding (Figure \ref{fig:saliency-bad}) using automatic differentiation \citep{Paszke2017}.
We compare against  LRFs that \citet{Kuang2018-wh} computed by perturbing the steady forcing of a CRM, which presumably represent causal relationships between the inputs and outputs.

Are the LRFs of our NNs similar to those of \citet[Figure 1]{Kuang2018-wh}?
The LRFs of \citet{Kuang2018-wh} have an upward dependence structure (a mainly upper-triangular structure in the panels of Figure \ref{fig:saliency-bad}), where the heating and to a lesser extent, the moistening, at a given vertical level depend only on data from below.  This is physically realistic since cumulus convection initiates at lower elevations and deepens into cumulonimbus; evaporation of precipitation leads to some low-triangular structure as well.   On the other hand, our NN predicts that heating and moistening at all heights are very sensitive to the humidity near the tropopause (200 mb), a physically nonsensical result.
A \SI{1}{\g\per\kg} increase of humidity at this level leads to more than \SI{10}{\g\per\kg\per\day} of drying at all lower levels.
Moreover, the NN does not accurately predict $Q_1$ and $Q_2$ at these levels (cf. Figure \ref{fig:r2}), perhaps because the fluctuations of humidity and temperature at these levels occur on faster time-scales than the 3-hour sampling rate of the data.
Thus, the NN depends most sensitively on the inputs that it performs worst on.

The NN is sensitive to upper-level humidity because the latter is strongly correlated to precipitation.
Figure \ref{fig:damping}a shows that the humidity and apparent moistening are highly anticorrelated, as measured by the effective damping rate given by
\begin{equation}
  \label{eq:damping}
  \gamma = -\frac{\langle Q_2 q_T \rangle}{\langle q_T q_T \rangle}
\end{equation}
where $\langle \cdot \rangle$ is the zonal and time averaging operator.
Near the tropopause, the moisture has a damping time-scale of \SI{3}{h}, which is the sampling rate of the NG-Aqua data.
This strong damping results in a near perfect correlation between the advective moistening and $q_T$ at this level, as shown in Figure \ref{fig:damping}b.

The advective moistening is primarily driven by grid-scale mean vertical motions due to cumulus convection.
The numerical instability probably appears in the tropics first because the weak temperature gradient there ensures that latent heating drives grid-scale average vertical motions \citep{Sobel2000-im}.
Thus, the amount of water cSAM advects into the upper atmosphere strongly correlates with precipitation.

The sensitivity to humidity can drive numerical instability as follows. 
A positive moisture anomaly in the tropical upper troposphere will increase the predicted precipitation and heating. 
This will then increase the upward velocity, which closes the feedback loop by increasing the supply of moisture to the upper atmosphere.
Ultimately, this feedback can lead to grid-scale storms (see Figure \ref{fig:snapshots}c below), similarly to moisture convergence closures.

This instability might be less problematic had we been able to customize the 4 km NG-Aqua simulations for the training process so as to better infer causality, e. g. by using higher output time resolution and calculation of instantaneous heating and moistening rates.   However, we were forced to work around it, as described in the next section.

\subsubsection{Eliminating non-causal relationships}
\label{sec:5.2}

We enforce a more plausible causal structure on the NN by not using the humidity and temperature above certain heights as predictors.
To obtain a numerically stable simulation without grid-scale storms, it suffices to ignore the $q_T$ and $s_L$ at or above levels 15 (\SI{450}{mb}) and 19 (\SI{225}{mb}), respectively.
We also assume that that the NN predicts $Q_1=Q_2=0$ at the same levels; however, a debiasing procedure discussed below will allow predictions of non-zero tendencies there.
To justify this assumption, we note that $Q_2$ is small at these levels, and that $Q_1$ is very noisy above \SI{250}{mb} (cf. Figure \ref{fig:q1q2}).
We have run cSAM simulations with this configuration for up to 100 days without
numerical blow-up or grid-point storms.

Figure \ref{fig:salience-good} shows the LRF of the modified NN.
Most strikingly, the linearized response of the drying/moistening to the input
variables is an order of magnitude smaller than for the NN with all
input variables.
For the most part, this LRF also has a smoother vertical structure that is more compatible with the CRM derived LRFs \citep{Kuang2018-wh}.
For instance, both our LRFs and those of \citet{Kuang2018-wh} show that destabilizing the atmosphere by decreasing the lower tropospheric temperature
leads to heating/drying throughout the column (panels b and d).
Likewise, the lower tropospheric moisture induces convective heating above (panel c).
Unfortunately, there are still large oscillations in the response of $Q_2$ to the humidity at \SI{500}{mb}.
Nonetheless, ignoring the upper atmospheric variables markedly improves the linearized response of the NN.

\subsection{Network architecture and training}
\label{sec:nn-details}

Unlike \citet{Rasp2018-ff}, we did not find that changing network architecture by adding more layers or parameters prevents blow-up. 
However, since we are training the data with global data, we use a higher capacity
network than \citet{Brenowitz2018-td} used for the tropics alone.
Thus, each model trained in this paper has three densely connected hidden layers
with 256 nodes each and rectified linear unit (ReLU) activations.
Altogether, this architecture has \num{160266} free parameters. 

Prior to input, each input variable is normalized by the global mean and
dividing by the standard deviation for each level independently, except where the latter vanishes.
This level-by-level normalization is a departure from past studies \citep{Rasp2018-ff,Brenowitz2018-td}, but had little impact upon the results.
In particular, NNs trained with inputs normalized as \citet{Brenowitz2018-td} also learn to depend strongly on the upper atmospheric humidity and suffer from coupled numerical instabilities discussed above.

The networks weights and biases are optimized by stochastic gradient descent (SGD)
with the popular Adam \citep{Kingma2014-bp} optimizer with an initial learning rate of \num{0.005}.
If a sample is defined as an individual atmospheric column for a given horizontal location and time step, then there are \num{5242880} samples in the training dataset. 
These samples are randomly in horizontal space and formed into batches of 64
time series, each containing the full 80 days of data.
In practice, SGD trains an NN that captures the time-space variability in diabatic processes, but makes large ($\sim \SI{1}{mm/d}$) errors in the zonal and time mean.  
This occurs because of the optimization is highly non-convex and the variability of cloud-related processes is larger than its mean.
Therefore, after SGD converges, we use linear regression to de-bias the predicted heating and moistening for each height and latitude separately (see Text S1).
To avoid re-introducing the causality and numerical stability issues solved in Section \ref{sec:coupled-instability}, these regressions depend only the predictions of $Q_1$ and $Q_2$ as well as $s_L$ and $q_T$ at the given height and do not introduce new dependencies between vertical levels.

The NN training and linear regression are implemented in Python using PyTorch \citep{Paszke2017} and scikit-learn \citep{scikit-learn}, respectively. 
The debiased NN models are then coupled to cSAM using a python interface.
The predicted tendencies are disabled within 3 grid cells (\SI{480}{km}) of the boundary because the flow in these regions is greatly influenced by the no-normal flow boundary conditions, so the NN is inaccurate there (cf. Fig.  \ref{fig:r2}).

\section{Results}
\label{sec:results}
\begin{figure}
  \centering
  \includegraphics[width=.7\textwidth]{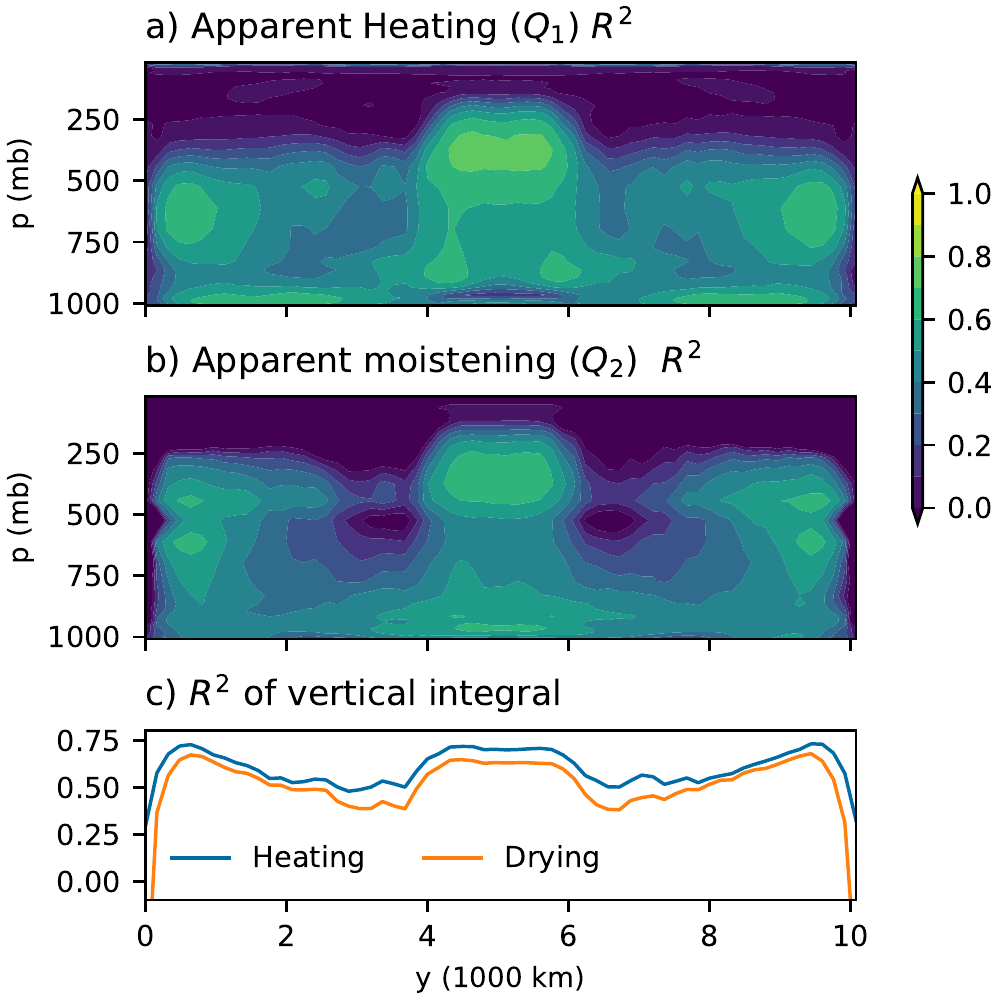}
  \caption{The $R^2$ of the $Q_1$ (a) and $Q_2$ (b) predicted by NN-lower, which ignores upper atmospheric inputs.
    Panel c shows the accuracy ($R^2$) for the mass-weighted vertical integrals
    of $Q_1$ (heating) and $Q_2$ (moistening).}

  \label{fig:r2}
\end{figure}

\begin{figure}
  \includegraphics[width=\textwidth]{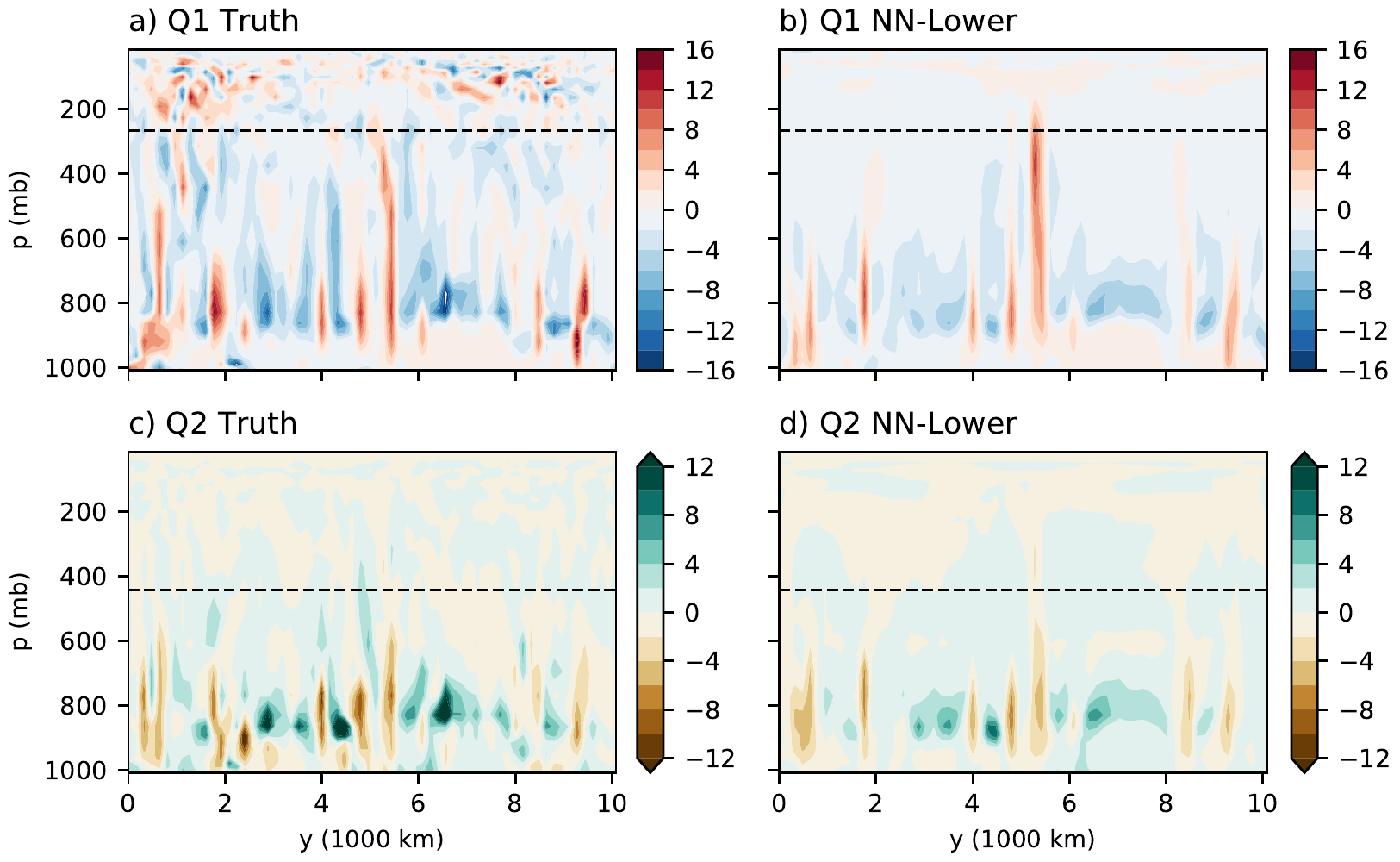}
  \caption{Comparison of the true $Q_1$ and $Q_2$ and instantaneous predictions with NN-Lower for $x=\SI{0}{km}$ at day 101.875. Dashed lines indicate the levels above which the NN predicts $Q_1= Q_2 =0 $. Note that the de-biasing linear regression model (Text S1) can predict a non-zero value there.}
  \label{fig:q1q2}
\end{figure}

\begin{figure}
  \includegraphics[width=\textwidth]{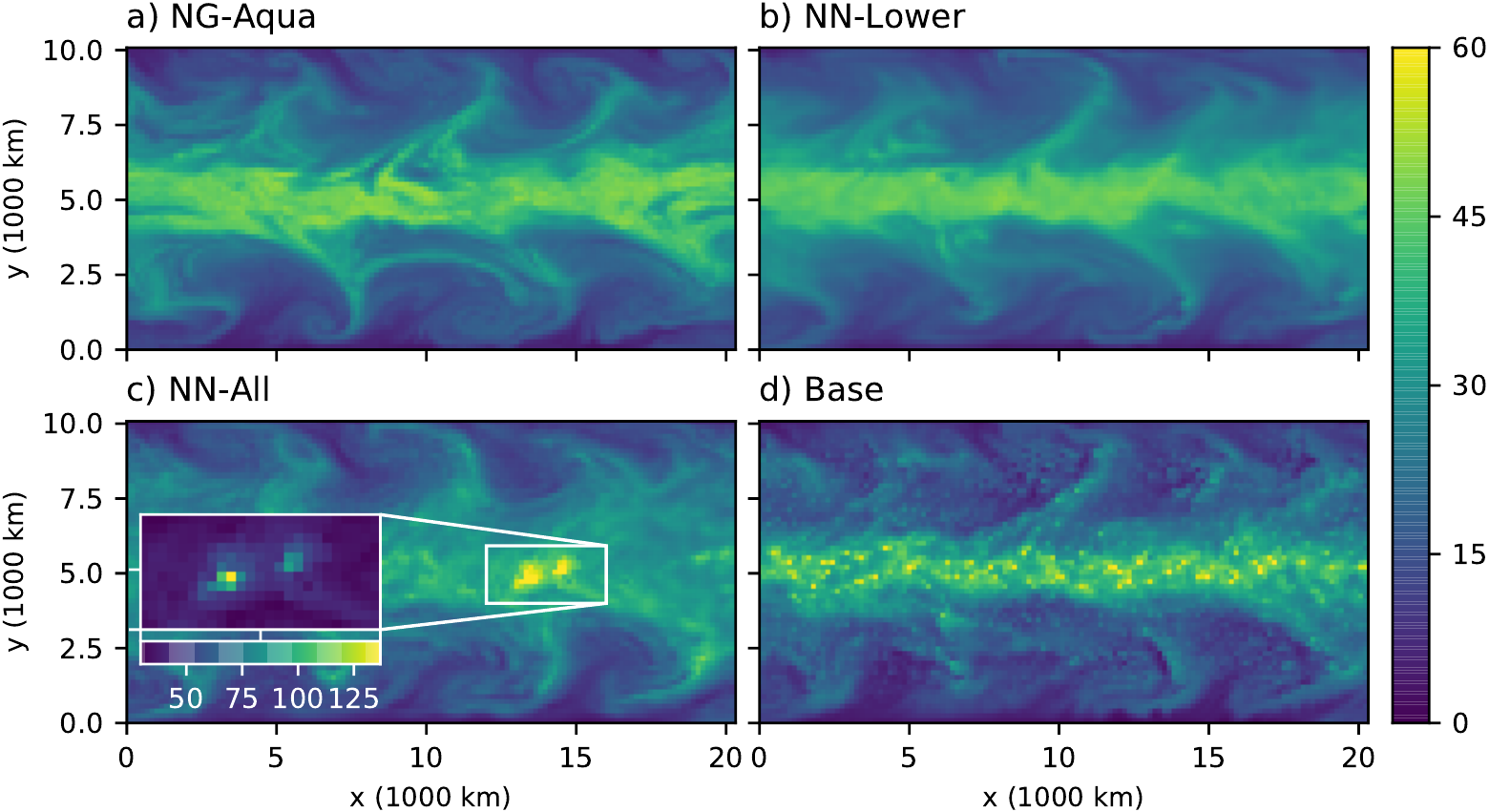}
  \caption{Precipitable water at day \num{105} for the truth (a) and cSAM
    simulations initialized at day \num{100.625} (b-c).  Grid-scale storms in NN-All (c)  saturate the color map, so a zoomed-in panel shows the full range of values. \label{fig:snapshots}}
\end{figure}

We now present our spatially-extended simulations.
As described in Section \ref{sec:gcm}, the simulations are initialized with the
NG-Aqua data at day 100.625, and our discussion focuses on the
prediction accuracy in the first 10 forecast days.
Two NN parametrizations will be compared with the training data
and the base simulation with resolved-scale microphysics and radiation. 
The first NN simulation, labeled as ``NN-All'', is performed using the NN
which includes all levels as input. 
The discussion in Section \ref{sec:coupled-instability} indicates that this
setup leads to grid-scale storms and model blow-up. 
The NN in the second simulation, known as ``NN-Lower'', only uses humidities below
level 16 (375 mb) and temperature below level 19 (226 mb) as described in
Section \ref{sec:5.2}.
When evaluated directly on the NG-Aqua data, NN-Lower has an
$R^2$ value of around 60\% for $Q_1$ and $Q_2$ in the lower troposphere (Figure
\ref{fig:r2}). 
Interestingly, the predicted heating and moistening are less noisy than the
``true'' values estimated as budget residuals (Figure \ref{fig:q1q2}).
Thus, the NN-Lower parametrization performs well in an
instantaneous sense.

Figure \ref{fig:snapshots}b demonstrates that the NN-Lower configuration is indeed
numerically stable and produces a reasonable prediction of the PW after 5 days,
albeit with a visible loss in large-scale moisture variability in the tropics. 
On the other hand, NN-All shows extremely moist grid-scale storms near the
tropics, which coincide with intense precipitation and ascent (not shown).
Likewise, the base simulation suffers from grid-point storms in the tropics
and frontal regions in the extratropics, and it ultimately blows up after 9 days of
simulation.

\subsection{Perfect Model Weather Prediction}

\begin{figure}
  \includegraphics[width=\textwidth]{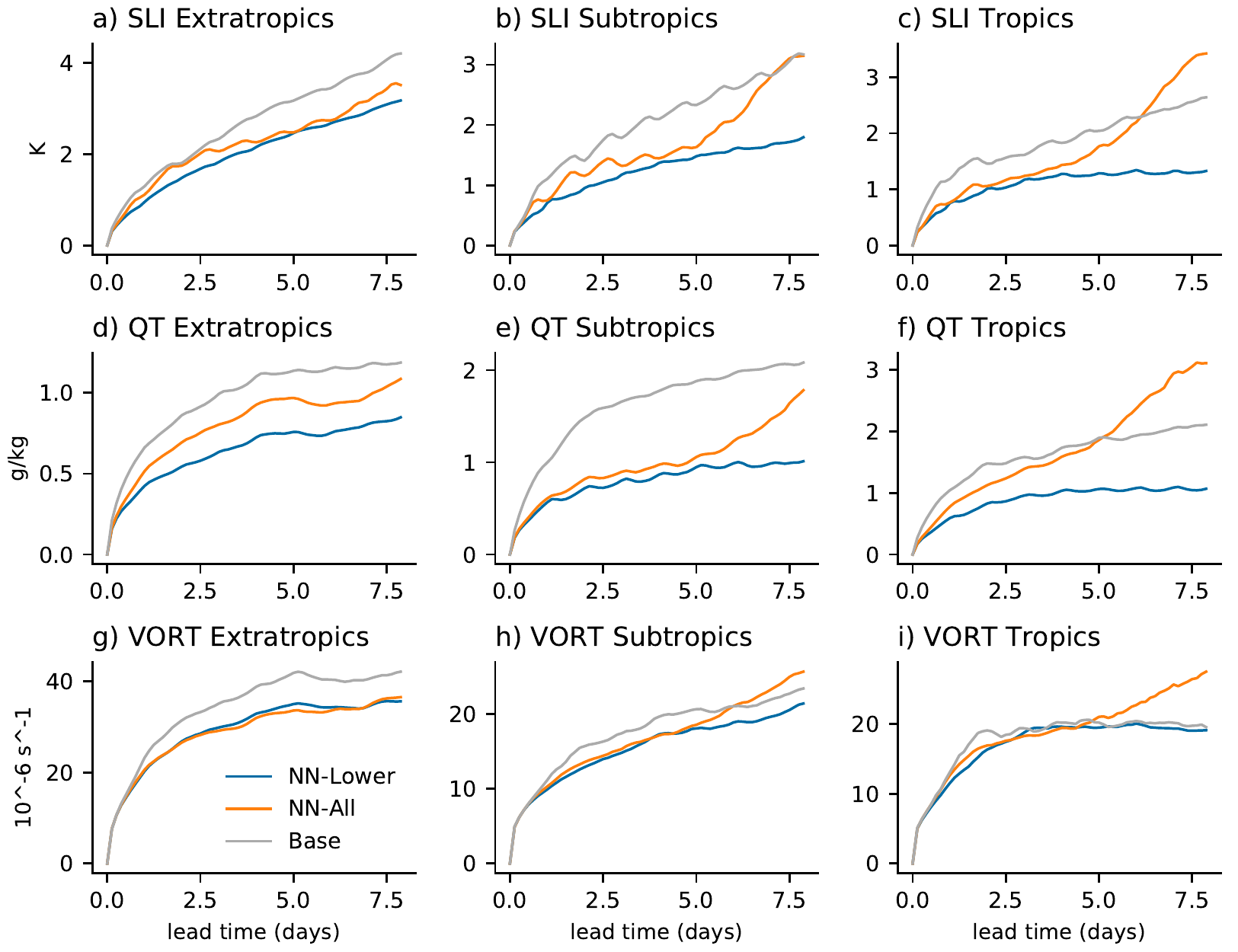}
  \caption{Root mean squared error (RMSE) evolution vs. forecast lead time for $s_L$ (a-c), $q_T$ (d-f), and
    vertical relative vorticity (g-i). 
    Separate curves are shown for each region (column) and cSAM
    configuration (indicated by color).}
\label{fig:rms}
\end{figure}

\begin{figure}
  \includegraphics{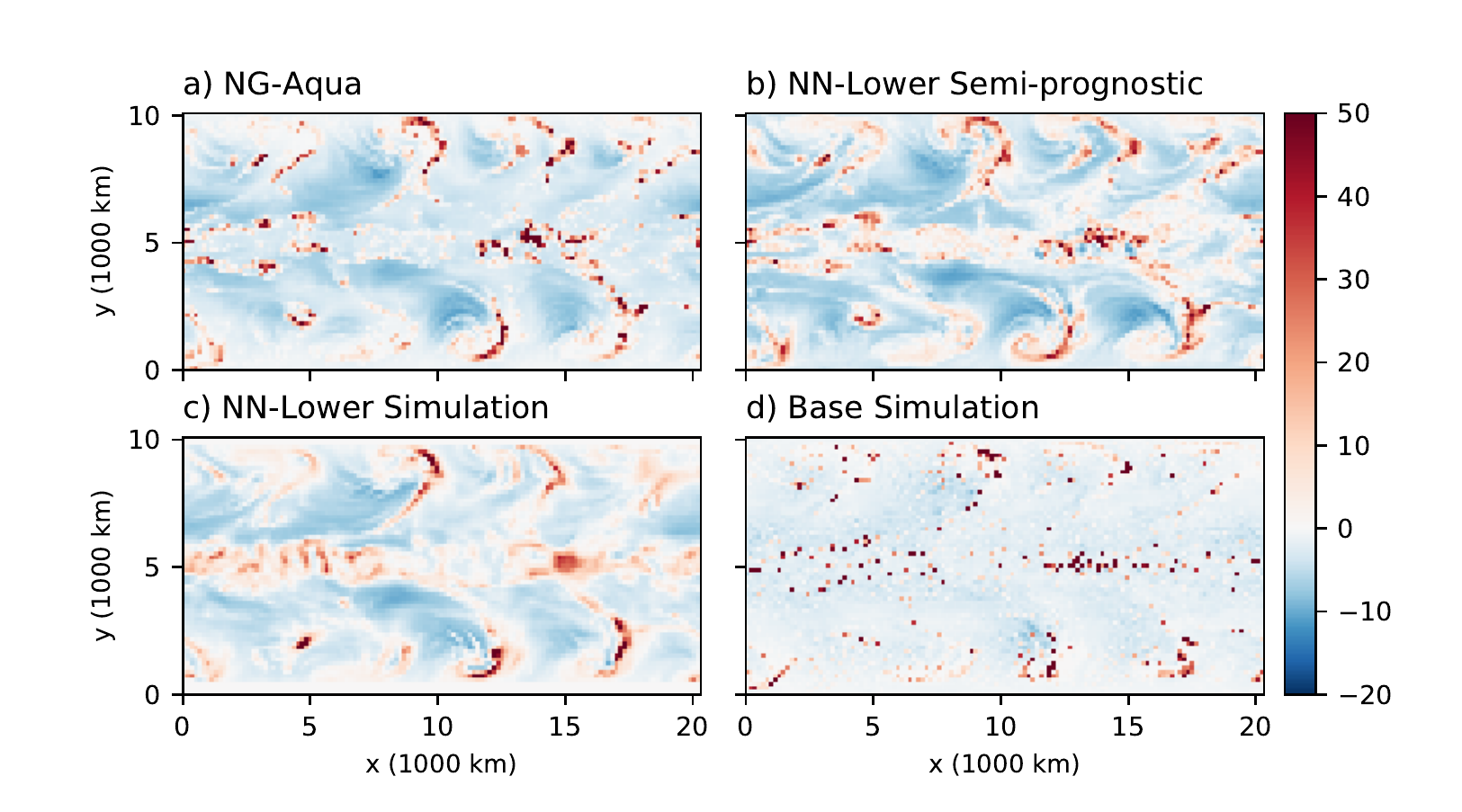}
  \caption{Comparison of forecasted and actual net precipitation after two forecast
    days ($\text{time}= \SI{102.625}{d}$).}
  \label{fig:map-precip}
\end{figure}

\begin{figure}
  \centering
  \includegraphics[width=\textwidth]{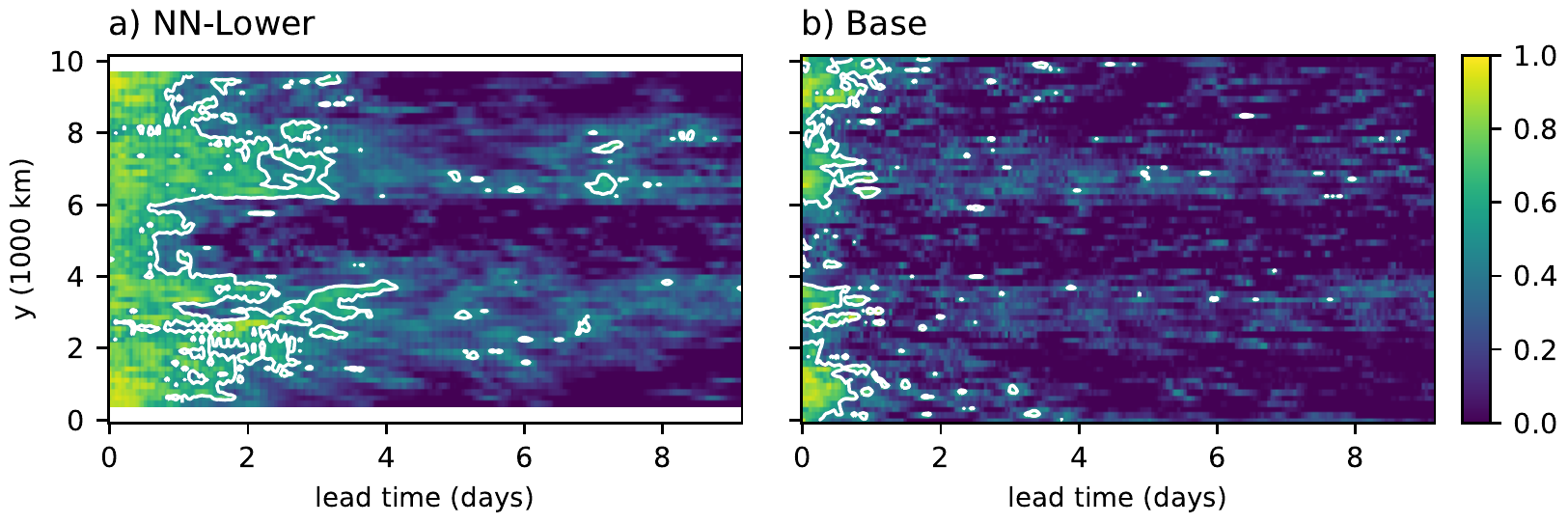}
  \caption{Hovmoller diagram of the net precipitation pattern correlation between the truth and
    the NN-Lower and base simulations. 
    The white contour indicates a correlation of 0.5.}
  \label{fig:precip-correlation}
\end{figure}

Figure \ref{fig:rms} shows the root-mean-squared error (RMSE) of $s_L$, $q_T$, and the
vertical component of vorticity for each configuration.
The RMSE values are vertically integrated and mass-weighted and averaged over 
different latitude bands.
The tropics lie within \SI{1280}{\km} of the equator, the sub-tropics are
between $y=\SI{1280}{\km}$ and \SI{2560}{\km}, and the extra-tropics are the poleward
\SI{2560}{\km} of the domain.
For all regions and variables, NN-Lower has the lowest RMSE.
This improvement is most striking for the thermodynamic variables $q_T$ and $s_L$, which have
parameterized source terms.
Compared to the base simulation, NN-Lower has 50\% lower errors in both $s_L$ and $q_T$.
The vorticity error improves less, which is not surprising since we do not attempt to correctly represent the source of momentum ($Q_3$).
However, in the extra-tropics, both NN-Lower and NN-All outperform the
base simulation.
Overall, NN-Lower produces the most accurate predictions. 

The NN-Lower simulation also predicts the net precipitation rate more accurately than the base simulation does.
The net precipitation rate is given by
\[ - \int \rho_0 Q_2 dz = P - E,\] where $P$ is precipitation and $E$ is the
evaporation \citep{Yanai1973-gt}. 
Net precipitation is the closest analog to precipitation that the NNs can predict
because they predict the combined effect of precipitation and evaporation.

The net precipitation is much less noisy than in NN-Lower than the base run.
Figure \ref{fig:map-precip} shows maps of net precipitation after two days of
prediction (day 102.625) for the NG-Aqua training data and the
NN-Lower and base simulations.
Overall, the NN-Lower simulation produces the correct extra-tropical pattern of
net precipitation, but fails to produce enough variability in the
tropics.
In NG-Aqua, the tropical precipitation occurs in small-scale clusters and in a
larger-scale Kelvin wave centered around $x=\SI{15000}{\km}$ at day 102.625, elsewhere there is column moistening due to evaporation.
On the other hand, the base simulation produces far too strong and noisy
precipitation, as expected for a scheme which only rains when a large-scale grid-cell becomes saturated.

Does the NN predict overly smooth net precipitation at the initial time or does this smoothness only appear as the coupled simulation evolves?
The NN's ``semi-prognostic'' net precipitation (i.e. predicted from the true coarse-grained NG-Aqua state at that time) is also shown in Figure \ref{fig:map-precip}b. 
Even in this instantaneous sense, the NN-Lower prefers to lightly rain throughout the tropics, rather than moisten in certain regions and strongly dry in others.

Figure \ref{fig:precip-correlation} shows the pattern correlation compared to
NG-Aqua of net precipitation for the NN-Lower and base simulations.
NN-Lower has higher pattern correlations than the base simulation at all times and retains a correlation of 0.5 up to day 101.5 in the tropics and beyond day 102.5 in the extra-tropics and sub-tropics.
By this measure, the base simulation has little predictive skill in the
tropics.

\subsection{Mean-state bias}

\begin{figure}
  \includegraphics{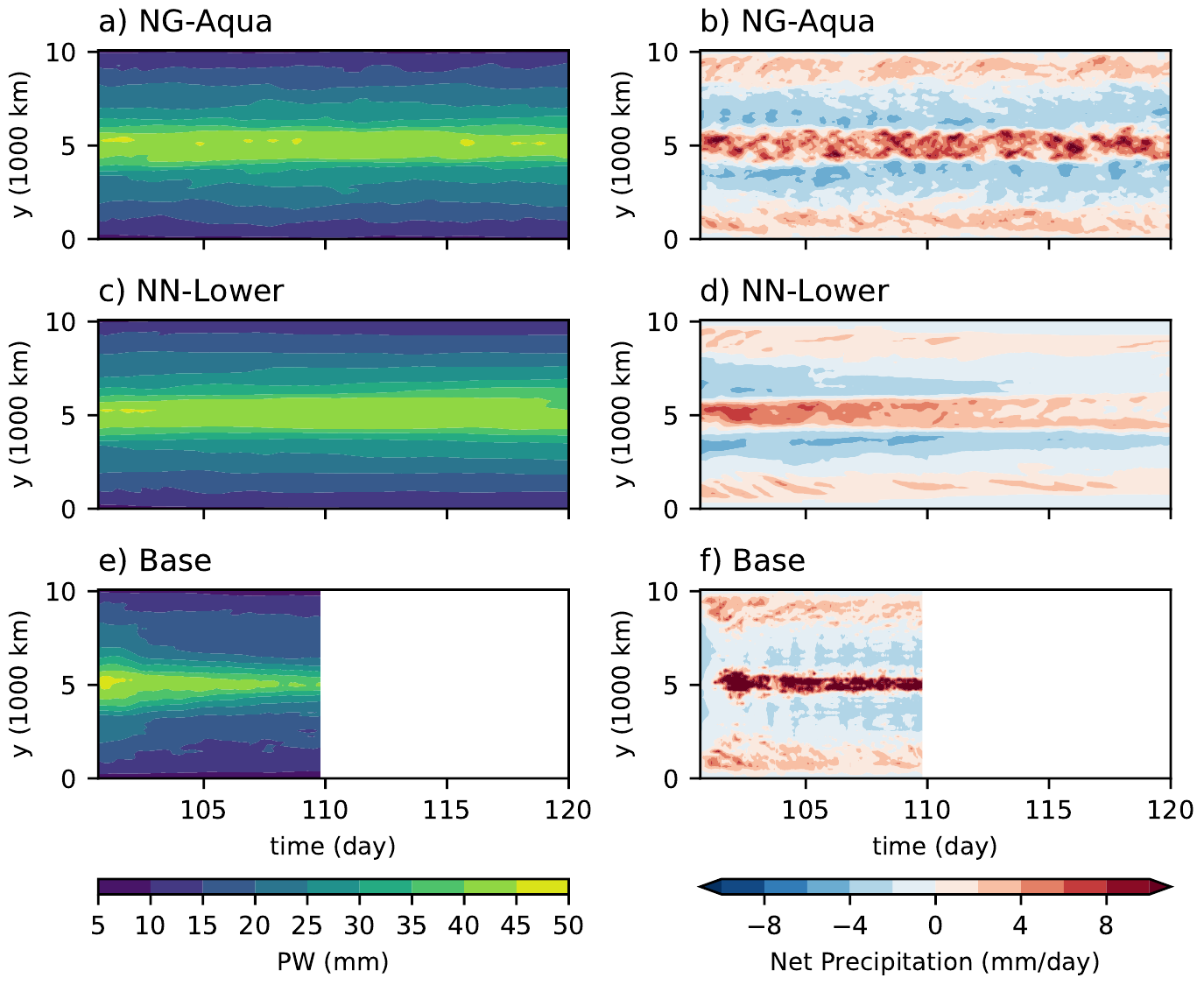}
  \caption{Hovmoller diagrams of the zonal mean of precipitable water (a, c, e), and
    net precipitation (b, d, f) for the training data (a, b),  NN-Lower (c, d), and
    base (e, f) simulations. White indicates that the simulation has blown-up.}
  \label{fig:zonal-means}
\end{figure}

\begin{figure}
  \includegraphics[width=\textwidth]{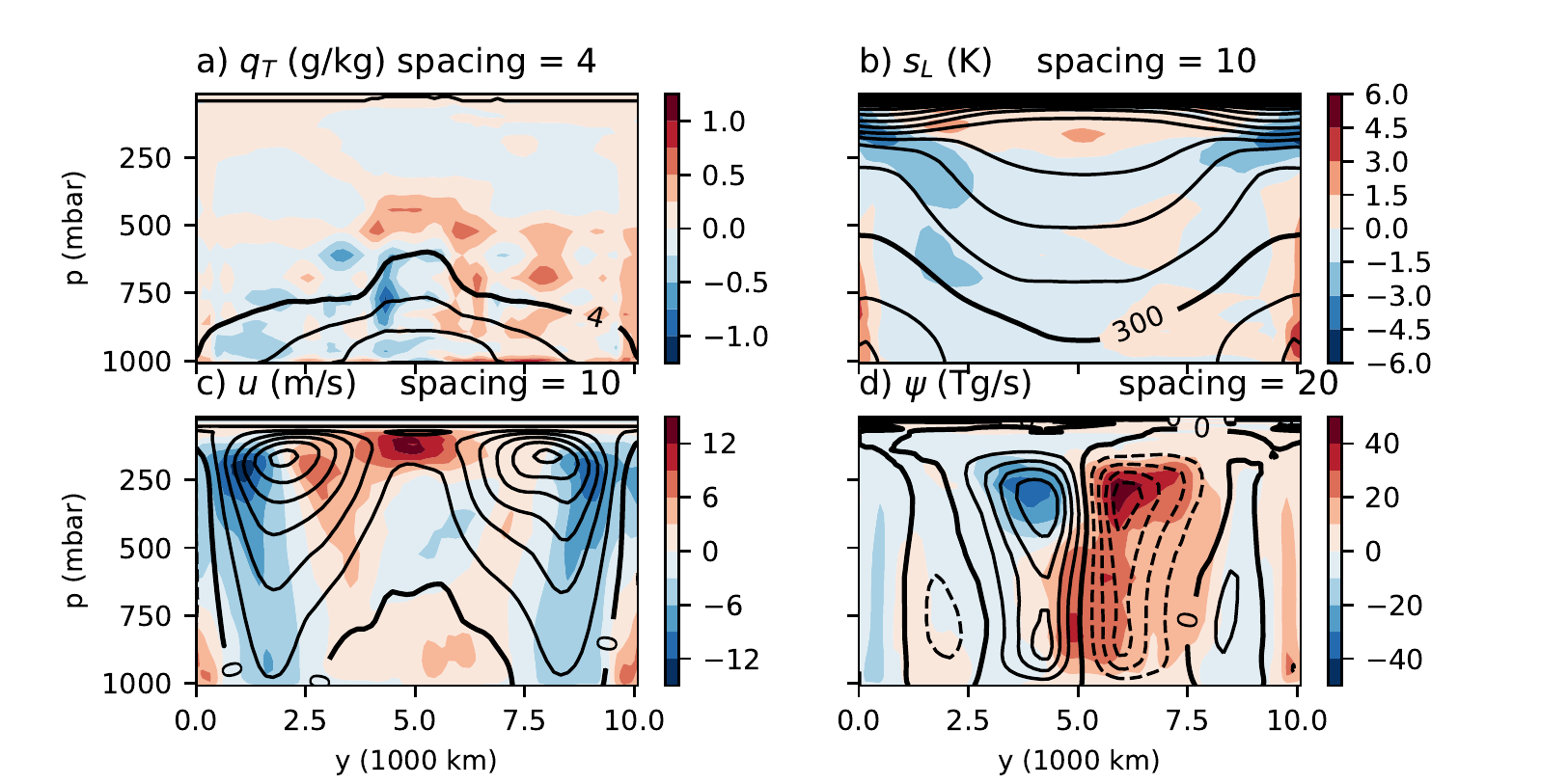}
  \caption{Zonal averages for $q_T$ (a), $s_L$ (b), $u$ (c), and streamfunction
    $\psi$ (d) for the final five forecast days. 
    The true climate (contours) and the bias of NN-Lower (colors) are overlayed.}
  \label{fig:bias}
\end{figure}

Short-term predictions with NN-Lower are reasonably accurate, but the mean state drifts away from the true climate. 
We first focus on the time evolution of the zonal mean precipitable water (PW)
and net precipitation, as shown in Figure \ref{fig:zonal-means}.
The fields are shown for days 100-120 for NG-Aqua and the NN-Lower simulation; the
base simulation crashed in day 110.
Compared to NG-Aqua, NN-Lower maintains the correct meridional
distribution of PW, but the net precipitation steadily decreases in the tropics.
All else equal, this would tend to increase the PW in the tropics, but the
Hadley circulation and its transport of vapor towards the equator also weaken.
On the other hand, the net precipitation in the base simulation becomes too
concentrated at the equator and atmosphere dries out significantly.
Overall, NN-Lower outperforms the base case and maintains the correct
zonal-mean PW distribution, but does not produce enough precipitation in the
tropics at later times.

Figure \ref{fig:bias} shows the zonal-mean bias vs. NG-Aqua for the zonal winds, meridional streamfunction, humidity \qt, and temperature \sli in the NN-Lower  simulation, time-averaged over forecast days 5--10. 
The humidity bias is relatively small with an amplitude of less than
\SI{1}{\g\per\kg}. 
The bias of \sli is also small, and peaks near the meridional boundaries of the domain where the NN is not used. 
On the other hand, the meridional circulation substantially weakens from a peak
mass transport of \SI{80e12}{\kg\per\s} to \SI{40e12}{\kg\per\s}.
This is associated with the decreasing $P-E$ in the tropics seen in
Figure \ref{fig:zonal-means}.
The zonal mean zonal winds also differ substantially. 
The surface westerlies and eddy-driven-jets become weaker, likely due to the
Held-Suarez momentum damping. 
At the same time, the jets shift southward and super-rotation develops near \SI{250}{mb}, likely as a transient response to the weakening Hadley circulation.
Overall, the small biases in the thermodynamic variables \qt and $s_L$ belie
growing biases in the circulation.

\begin{figure}
  \includegraphics{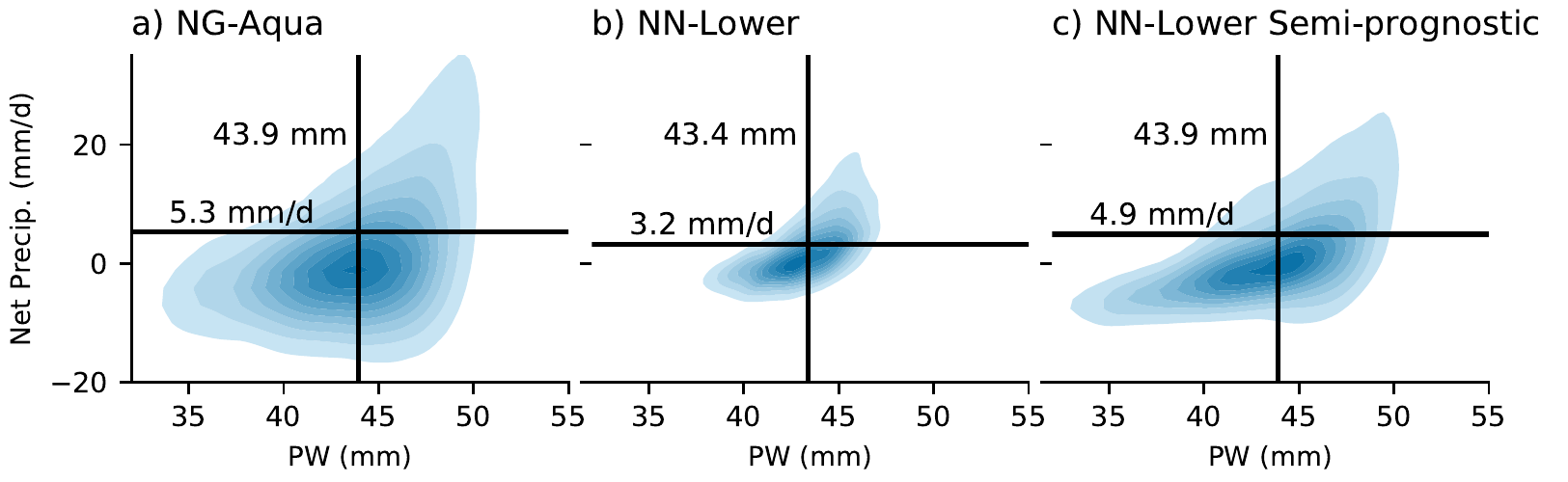}
  \caption{Joint probability distribution functions of PW and net precipitation for  the tropics estimated for days 110--112. The means of PW and net precipitation are indicated with text and black lines.}
  \label{fig:pdf}
\end{figure}

The loss of zonal variability in net precipitation (Fig. 
\ref{fig:map-precip}) and precipitable water (Fig. 
\ref{fig:snapshots}) might explain these mean-state drifts.
Precipitation depends exponentially on precipitable water \citep{Rushley2018-me,Bretherton2004-jf}, therefore a reduction in the zonal variance of PW (fewer extremely moist
  or dry coarse grid cells) will tend to
decreases the zonal-mean precipitation for a given zonal-mean PW.
Figure \ref{fig:pdf} confirms that this occurs in the NN-Lower simulation, whose mean tropical net precipitation is 40\% weaker than NG-Aqua despite having a similar mean PW.
Indeed, Figure \ref{fig:pdf} shows that the joint distribution of PW and net precipitation has much less variance in NN-Lower than in NG-Aqua.

The predicted net precipitation has too little variance even when evaluated on the true state (Fig. \ref{fig:pdf}c).
In other words, the distribution of net precipitation has too little spread for a given value of PW--- i.e. the NN relies too strongly on PW.
When integrated forward in time, this lack of $P-E$ variance likely reduces the PW variance, and therefore the mean precipitation.
Thus, enhancing the variance of the conditional distributions of net precipitation could ameliorate the circulation biases in these simulations.

Plausible physical mechanisms for the loss of moisture variability depend on the spatial scale.
On smaller scales, both the explicit hyper-diffusion and implicit numerical
diffusion could reduce the variance of moisture.
For larger scales, the moisture probably varies less because NN-Lower
prefers to smooth $P-E$ throughout the tropics rather than delineating precipitating and
evaporating regions.
Future work could improve this by developing a stochastic or deterministic trigger for convection, or by separating precipitating and evaporating processes.

\section{Conclusions}
\label{sec:conclusion}

We use a neural network to build a unified physics parametrization of diabatic heating and moistening, which include the effects of unresolved subgrid processes.
The neural network is trained by coarse-graining a near global aquaplanet simulation with a grid size of \SI{4}{km} to a resolution of \SI{160}{km}.
It learns the coarse-grid heating and moistening profiles, calculated as residuals from the coarse-grid dynamics, from the temperature and moisture profiles, plus auxiliary parameters. 
A key challenge of the coarse-graining approach
is that it lacks the natural hierarchical structure of other training datasets used in
the literature, such as super-parametrization (SP) \citep{Rasp2018-ff}.  
In the SP application, for instance, the target model and the training model share the same software interface for the physical process tendencies.  These tendencies are predicted by a high-resolution submodel in the SP training dataset and must be learned by the neural network parametrization for application to the target model.   This application is thus a clear target for emulation.

We extend the single column model results of \citet{Brenowitz2018-td} to a
spatially-extended simulation.
The coarse-grid dynamical core is the same anelastic atmospheric model that generated the
4~km training dataset but is run with a \SI{160}{km} resolution. 
This model needed additional damping in the form of horizontal hyper-diffusion
and Newtonian relaxation near the surface to run stably in a base
configuration with its default resolved-scale microphysics and radiation schemes, or with our neural network unified-physics parametrization.  We developed a deeper version of our neural network capable of simulating the diabatic
heating and moistening over the entire 46S-46N domain of our training simulations, rather than just the tropical subdomain used by \citet{Brenowitz2018-td}. We include a zonal-mean bias correction to minimize zonal mean temperature and moisture drifts.

An attempt to couple a preliminary version of this neural network to this GCM caused the model to blow up.
Analyzing the linearized response of that neural network showed that it inadvertently learns to exploit a strong correlation between upper atmospheric humidity and precipitation.
This correlation owes to the short lifetime of water vapor at those
heights rather than any causal mechanism.
Thus, ML parametrization is not immune to issues with causality that have long
inspired debates about closure assumptions.

We enforced a plausible causal structure by removing the upper atmospheric
humidity and temperature from the neural network inputs.
Spatially extended simulations with this modified network can run stably indefinitely, without blowing up.
Thus, stabilizing the parametrization required a rather crude human
intervention.
Future studies will need to explore automatic ways to discover true causal
relationships and forestall model blow-up in a dynamically coupled setting.

The resulting neural network predicted the weather of NG-Aqua with much higher forecast skill and lower bias than a base coarse-grid simulation with only resolved microphysics and radiation parametrizations, which is the only reference case we could easily use as a metric.
For a fair comparison, we will need to train and implement the ML parametrization in a
global model whose coarse-grid version is typically run with a traditional suite of physical parametrizations.

More work is needed to reduce the long-term biases in these simulations.
While temperature and humidity biases are small in the first 10 days, the Hadley
circulation is dramatically weakened because there is not enough heating in the
tropics to sustain vertical motions. 
Thus, more work is needed to keep the climate generated by neural network parametrizations trained by coarse-graining from drifting.
We argue that our scheme could benefit from adding a mechanism for enhancing the tropical variance of moisture and the predicted precipitation, such as stochasticity in the parametrized tendencies.

\acknowledgments

N.B. is supported as a postdoctoral fellow by the Washington Research Foundation, and by a Data Science Environments project award from the Gordon and Betty Moore Foundation (Award \#2013-10-29) and the Alfred P. Sloan Foundation (Award \#3835) to the University of Washington eScience Institute. C.B. is supported by U. S. Department of Energy grants DE-SC0012451 and DE-SC0016433.
The coarse-grained NG-Aqua data and computer codes are available at \verb|zenodo.org| \citep{ngaqua-data-2018,ngaqua-data} and GitHub \citep{github}, respectively.







%
%
%
%
%
%
%
%
%
%

\bibliography{main}

\begin{thebibliography}{46}
\providecommand{\natexlab}[1]{#1}
\expandafter\ifx\csname urlstyle\endcsname\relax
  \providecommand{\doi}[1]{doi:\discretionary{}{}{}#1}\else
  \providecommand{\doi}{doi:\discretionary{}{}{}\begingroup
  \urlstyle{rm}\Url}\fi

\bibitem[{\textit{Arakawa}(2004)}]{Arakawa2004-io}
Arakawa, A. (2004), The cumulus parameterization problem: Past, present, and
  future, \textit{J. Clim.}, \textit{17}(13), 2493--2525,
  \doi{10.1175/1520-0442(2004)017<2493:RATCPP>2.0.CO;2}.

\bibitem[{\textit{Brenowitz}(2018)}]{ngaqua-data-2018}
Brenowitz, N.~D. (2018), Coarse-grained outputs from near-global aqua-planet
  control run with {QOBS} {SST}, \doi{10.5281/zenodo.1226370}.

\bibitem[{\textit{Brenowitz}(2019{\natexlab{a}})}]{ngaqua-data}
Brenowitz, N.~D. (2019{\natexlab{a}}), Coarse-grained near-global aqua-planet
  simulation with computed dynamical tendencies, \doi{10.5281/zenodo.2621638}.

\bibitem[{\textit{Brenowitz}(2019{\natexlab{b}})}]{github}
Brenowitz, N.~D. (2019{\natexlab{b}}), \verb|https://github.com/nbren12/uwnet|
  v0.7, \doi{10.5281/zenodo.2621717}.

\bibitem[{\textit{Brenowitz and Bretherton}(2018)}]{Brenowitz2018-td}
Brenowitz, N.~D., and C.~S. Bretherton (2018), Prognostic validation of a
  neural network unified physics parameterization, \textit{Geophys. Res.
  Lett.}, \textit{17}, 2493, \doi{10.1029/2018GL078510}.

\bibitem[{\textit{Bretherton and Khairoutdinov}(2015)}]{Bretherton2015-iz}
Bretherton, C.~S., and M.~F. Khairoutdinov (2015), Convective self-aggregation
  feedbacks in near-global cloud-resolving simulations of an aquaplanet,
  \textit{Journal of Advances in Modeling Earth Systems}, \textit{7}(4),
  1765--1787.

\bibitem[{\textit{Bretherton et~al.}(2004)\textit{Bretherton, Peters, and
  Back}}]{Bretherton2004-jf}
Bretherton, C.~S., M.~E. Peters, and L.~E. Back (2004), Relationships between
  water vapor path and precipitation over the tropical oceans, \textit{J.
  Clim.}, \textit{17}(7), 1517--1528,
  \doi{10.1175/1520-0442(2004)017<1517:RBWVPA>2.0.CO;2}.

\bibitem[{\textit{Chevallier et~al.}(1998)\textit{Chevallier, Ch{\'e}ruy,
  Scott, and Ch{\'e}din}}]{Chevallier1998-su}
Chevallier, F., F.~Ch{\'e}ruy, N.~A. Scott, and A.~Ch{\'e}din (1998), A neural
  network approach for a fast and accurate computation of a longwave radiative
  budget, \textit{J. Appl. Meteorol.}, \textit{37}(11), 1385--1397,
  \doi{10.1175/1520-0450(1998)037<1385:ANNAFA>2.0.CO;2}.

\bibitem[{\textit{Cybenko}(1989)}]{Cybenko1989-hr}
Cybenko, G. (1989), Approximation by superpositions of a sigmoidal function,
  \textit{Math. Control Signals Systems}, \textit{2}(4), 303--314,
  \doi{10.1007/BF02551274}.

\bibitem[{\textit{ECMWF}(2018)}]{ECMWF2018}
ECMWF (Ed.) (2018), \textit{Part III : Dynamics and numerical procedures},
  no.~3 in IFS Documentation, ECMWF.

\bibitem[{\textit{Emanuel et~al.}(1994)\textit{Emanuel, David~Neelin, and
  Bretherton}}]{Emanuel1994-fc}
Emanuel, K.~A., J.~David~Neelin, and C.~S. Bretherton (1994), On large-scale
  circulations in convecting atmospheres, \textit{Q.J.R. Meteorol. Soc.},
  \textit{120}(519), 1111--1143, \doi{10.1002/qj.49712051902}.

\bibitem[{\textit{Goodfellow et~al.}(2016)\textit{Goodfellow, Bengio, and
  Courville}}]{Goodfellow2016-wf}
Goodfellow, I., Y.~Bengio, and A.~Courville (2016), \textit{Deep learning},
  Adaptive computation and machine learning, The MIT Press, Cambridge,
  Massachusetts.

\bibitem[{\textit{Hanin}(2017)}]{Hanin2017-hu}
Hanin, B. (2017), Universal function approximation by deep neural nets with
  bounded width and {ReLU} activations, \textit{arXiv e-prints}.

\bibitem[{\textit{Held and Suarez}(1994)}]{Held1994-py}
Held, I.~M., and M.~J. Suarez (1994), A proposal for the intercomparison of the
  dynamical cores of atmospheric general circulation models, \textit{Bull. Am.
  Meteorol. Soc.}, \textit{75}(10), 1825--1830,
  \doi{10.1175/1520-0477(1994)075<1825:APFTIO>2.0.CO;2}.

\bibitem[{\textit{Jiang}(2017)}]{Jiang2017-sr}
Jiang, X. (2017), Key processes for the eastward propagation of the
  {Madden-Julian} oscillation based on multimodel simulations, \textit{J.
  Geophys. Res. D: Atmos.}, \textit{122}(2), 2016JD025,955,
  \doi{10.1002/2016JD025955}.

\bibitem[{\textit{Jiang et~al.}(2015)\textit{Jiang, Waliser, Xavier, Petch,
  Klingaman, Woolnough, Guan, Bellon, Crueger, DeMott, Hannay, Lin, Hu, Kim,
  Lappen, Lu, Ma, Miyakawa, Ridout, Schubert, Scinocca, Seo, Shindo, Song,
  Stan, Tseng, Wang, Wu, Wu, Wyser, Zhang, and Zhu}}]{Jiang2015-jo}
Jiang, X., D.~E. Waliser, P.~K. Xavier, J.~Petch, N.~P. Klingaman, S.~J.
  Woolnough, B.~Guan, G.~Bellon, T.~Crueger, C.~DeMott, C.~Hannay, H.~Lin,
  W.~Hu, D.~Kim, C.-L. Lappen, M.-M. Lu, H.-Y. Ma, T.~Miyakawa, J.~A. Ridout,
  S.~D. Schubert, J.~Scinocca, K.-H. Seo, E.~Shindo, X.~Song, C.~Stan, W.-L.
  Tseng, W.~Wang, T.~Wu, X.~Wu, K.~Wyser, G.~J. Zhang, and H.~Zhu (2015),
  Vertical structure and physical processes of the {Madden-Julian} oscillation:
  Exploring key model physics in climate simulations, \textit{J. Geophys. Res.
  D: Atmos.}, \textit{120}(10), 2014JD022,375, \doi{10.1002/2014JD022375}.

\bibitem[{\textit{Khairoutdinov and Randall}(2003)}]{Khairoutdinov2003-du}
Khairoutdinov, M.~F., and D.~A. Randall (2003), Cloud resolving modeling of the
  {ARM} summer 1997 {IOP}: Model formulation, results, uncertainties, and
  sensitivities, \textit{J. Atmos. Sci.}, \textit{60}(4), 607--625,
  \doi{10.1175/1520-0469(2003)060<0607:CRMOTA>2.0.CO;2}.

\bibitem[{\textit{Kim et~al.}(2011)\textit{Kim, Jang, Kim, Kim, Watanabe, Jin,
  and Kug}}]{Kim2011-de}
Kim, D., Y.-S. Jang, D.-H. Kim, Y.-H. Kim, M.~Watanabe, F.-F. Jin, and J.-S.
  Kug (2011), El ni{\~n}o-southern oscillation sensitivity to cumulus
  entrainment in a coupled general circulation model: {ENSO} {SENSITIVE} {TO}
  {CONVECTION} {SCHEME}, \textit{J. Geophys. Res.}, \textit{116}(D22),
  \doi{10.1029/2011JD016526}.

\bibitem[{\textit{Kingma and Ba}(2014)}]{Kingma2014-bp}
Kingma, D.~P., and J.~Ba (2014), Adam: A method for stochastic optimization.

\bibitem[{\textit{Krasnopolsky et~al.}(2005)\textit{Krasnopolsky,
  Fox-Rabinovitz, and Chalikov}}]{Krasnopolsky2005-ca}
Krasnopolsky, V.~M., M.~S. Fox-Rabinovitz, and D.~V. Chalikov (2005), New
  approach to calculation of atmospheric model physics: Accurate and fast
  neural network emulation of longwave radiation in a climate model,
  \textit{Mon. Weather Rev.}, \textit{133}(5), 1370--1383,
  \doi{10.1175/MWR2923.1}.

\bibitem[{\textit{Krasnopolsky et~al.}(2010)\textit{Krasnopolsky,
  Fox-Rabinovitz, and Belochitski}}]{Krasnopolsky2010-nn}
Krasnopolsky, V.~M., M.~S. Fox-Rabinovitz, and A.~A. Belochitski (2010),
  Development of neural network convection parameterizations for numerical
  climate and weather prediction models using cloud resolving model
  simulations, in \textit{The 2010 International Joint Conference on Neural
  Networks ({IJCNN})}, pp. 1--8, \doi{10.1109/IJCNN.2010.5596766}.

\bibitem[{\textit{Krasnopolsky et~al.}(2013)\textit{Krasnopolsky,
  Fox-Rabinovitz, and Belochitski}}]{Krasnopolsky2013-zw}
Krasnopolsky, V.~M., M.~S. Fox-Rabinovitz, and A.~A. Belochitski (2013), Using
  ensemble of neural networks to learn stochastic convection parameterizations
  for climate and numerical weather prediction models from data simulated by a
  cloud resolving model, \textit{Advances in Artificial Neural Systems},
  \textit{2013}, e485,913, \doi{10.1155/2013/485913}.

\bibitem[{\textit{Kuang}(2018)}]{Kuang2018-wh}
Kuang, Z. (2018), Linear stability of moist convecting atmospheres part i: from
  linear response functions to a simple model and applications to convectively
  coupled waves, \textit{J. Atmos. Sci.}, \doi{10.1175/JAS-D-18-0092.1}.

\bibitem[{\textit{Langenbrunner and Neelin}(2017)}]{Langenbrunner2017-ed}
Langenbrunner, B., and J.~D. Neelin (2017), {Pareto-Optimal} estimates of
  california precipitation change, \textit{Geophys. Res. Lett.}, p.
  2017GL075226, \doi{10.1002/2017GL075226}.

\bibitem[{\textit{Lyu et~al.}()\textit{Lyu, K{\"o}hl, Matei, and
  Stammer}}]{Lyu_undated-qw}
Lyu, G., A.~K{\"o}hl, I.~Matei, and D.~Stammer (), {Adjoint-Based} climate
  model tuning: Application to the planet simulator, \textit{J. Adv. Model.
  Earth Syst.}, \doi{10.1002/2017MS001194}.

\bibitem[{\textit{Majda}(2007)}]{Majda2007-qr}
Majda, A.~J. (2007), Multiscale models with moisture and systematic strategies
  for superparameterization, \textit{J. Atmos. Sci.}, \textit{64}(7),
  2726--2734, \doi{10.1175/JAS3976.1}.

\bibitem[{\textit{Mapes and Neale}(2011)}]{Mapes2011-vh}
Mapes, B., and R.~Neale (2011), Parameterizing convective organization to
  escape the entrainment dilemma, \textit{J. Adv. Model. Earth Syst.},
  \textit{3}(2), M06,004, \doi{10.1029/2011MS000042}.

\bibitem[{\textit{Nakazawa}(1988)}]{Nakazawa1988-xh}
Nakazawa, T. (1988), Tropical super clusters within intraseasonal variations
  over the western pacific, \textit{Journal of the Meteorological Society of
  Japan. Ser. II}, \textit{66}(6), 823--839.

\bibitem[{\textit{Narenpitak et~al.}(2017)\textit{Narenpitak, Bretherton, and
  Khairoutdinov}}]{Narenpitak2017-ep}
Narenpitak, P., C.~S. Bretherton, and M.~F. Khairoutdinov (2017), Cloud and
  circulation feedbacks in a near-global aquaplanet cloud-resolving model:
  Cloud feedbacks in a {Near-Global} {CRM}, \textit{J. Adv. Model. Earth
  Syst.}, \textit{9}(2), 1069--1090, \doi{10.1002/2016MS000872}.

\bibitem[{\textit{NOAA}(2018)}]{NOAA2018}
NOAA (2018), Strategic implementation plan for evolution of nggps to a national
  unified modeling system, \textit{Tech. rep.}

\bibitem[{\textit{O'Gorman and Dwyer}(2018)}]{OGorman2018-hn}
O'Gorman, P.~A., and J.~G. Dwyer (2018), Using machine learning to parameterize
  moist convection: Potential for modeling of climate, climate change, and
  extreme events, \textit{J. Adv. Model. Earth Syst.}, \textit{10}(10),
  2548--2563, \doi{10.1029/2018MS001351}.

\bibitem[{\textit{Palmer}(2001)}]{Palmer2001-rg}
Palmer, T.~N. (2001), A nonlinear dynamical perspective on model error: A
  proposal for non-local stochastic-dynamic parametrization in weather and
  climate prediction models, \textit{Quart. J. Roy. Meteor. Soc.},
  \textit{127}(572), 279--304, \doi{10.1002/qj.49712757202}.

\bibitem[{\textit{Paszke et~al.}(2017)\textit{Paszke, Gross, Chintala, Chanan,
  Yang, DeVito, Lin, Desmaison, Antiga, and Lerer}}]{Paszke2017}
Paszke, A., S.~Gross, S.~Chintala, G.~Chanan, E.~Yang, Z.~DeVito, Z.~Lin,
  A.~Desmaison, L.~Antiga, and A.~Lerer (2017), Automatic differentiation in
  {PyTorch}.

\bibitem[{\textit{Pecora et~al.}(1997)\textit{Pecora, Carroll, Johnson, Mar,
  and Heagy}}]{Pecora1997-kz}
Pecora, L.~M., T.~L. Carroll, G.~A. Johnson, D.~J. Mar, and J.~F. Heagy (1997),
  Fundamentals of synchronization in chaotic systems, concepts, and
  applications, \textit{Chaos}, \textit{7}(4), 520--543,
  \doi{10.1063/1.166278}.

\bibitem[{\textit{Pedregosa et~al.}(2011)\textit{Pedregosa, Varoquaux,
  Gramfort, Michel, Thirion, Grisel, Blondel, Prettenhofer, Weiss, Dubourg,
  Vanderplas, Passos, Cournapeau, Brucher, Perrot, and
  Duchesnay}}]{scikit-learn}
Pedregosa, F., G.~Varoquaux, A.~Gramfort, V.~Michel, B.~Thirion, O.~Grisel,
  M.~Blondel, P.~Prettenhofer, R.~Weiss, V.~Dubourg, J.~Vanderplas, A.~Passos,
  D.~Cournapeau, M.~Brucher, M.~Perrot, and E.~Duchesnay (2011), Scikit-learn:
  Machine learning in {P}ython, \textit{Journal of Machine Learning Research},
  \textit{12}, 2825--2830.

\bibitem[{\textit{Quarteroni et~al.}(2007)\textit{Quarteroni, Sacco, and
  Saleri}}]{Quarteroni2007-cq}
Quarteroni, A., R.~Sacco, and F.~Saleri (2007), \textit{Numerical Mathematics},
  Texts in Applied Mathematics, 2 ed., Springer-Verlag Berlin Heidelberg,
  \doi{10.1007/b98885}.

\bibitem[{\textit{Rasp et~al.}(2018)\textit{Rasp, Pritchard, and
  Gentine}}]{Rasp2018-ff}
Rasp, S., M.~S. Pritchard, and P.~Gentine (2018), Deep learning to represent
  subgrid processes in climate models, \textit{Proc. Natl. Acad. Sci. U. S.
  A.}, \textit{115}(39), 9684--9689, \doi{10.1073/pnas.1810286115}.

\bibitem[{\textit{Rushley et~al.}(2018)\textit{Rushley, Kim, Bretherton, and
  Ahn}}]{Rushley2018-me}
Rushley, S.~S., D.~Kim, C.~S. Bretherton, and M.-S. Ahn (2018), Reexamining the
  nonlinear {Moisture-Precipitation} relationship over the tropical oceans,
  \textit{Geophys. Res. Lett.}, \textit{45}(2), 2017GL076,296,
  \doi{10.1002/2017GL076296}.

\bibitem[{\textit{Satoh et~al.}(2008)\textit{Satoh, Matsuno, Tomita, Miura,
  Nasuno, and Iga}}]{Satoh2008-pt}
Satoh, M., T.~Matsuno, H.~Tomita, H.~Miura, T.~Nasuno, and S.~Iga (2008),
  Nonhydrostatic icosahedral atmospheric model ({NICAM}) for global cloud
  resolving simulations, \textit{J. Comput. Phys.}, \textit{227}(7),
  3486--3514, \doi{10.1016/j.jcp.2007.02.006}.

\bibitem[{\textit{Schneider et~al.}(2017)\textit{Schneider, Lan, Stuart, and
  Teixeira}}]{Schneider2017-qq}
Schneider, T., S.~Lan, A.~Stuart, and J.~Teixeira (2017), Earth system modeling
  2.0: A blueprint for models that learn from observations and targeted
  {High-Resolution} simulations, \textit{Geophys. Res. Lett.}, \textit{44}(24),
  12,396--12,417, \doi{10.1002/2017GL076101}.

\bibitem[{\textit{Sobel and Bretherton}(2000)}]{Sobel2000-im}
Sobel, A.~H., and C.~S. Bretherton (2000), Modeling tropical precipitation in a
  single column, \textit{J. Clim.}, \textit{13}(24), 4378--4392,
  \doi{10.1175/1520-0442(2000)013<4378:MTPIAS>2.0.CO;2}.

\bibitem[{\textit{Stevens et~al.}(Submitted)\textit{Stevens, Satoh, Auger,
  Biercamp, Bretherton, Chen, Dueben, Judt, Khairoutdinov, Klocke, Kodama,
  Kornblueh, Lin, Putman, Shibuya, Neumann, Roeber, Vanniere, Vidale, Wedi, and
  Zhou}}]{Stevens2019}
Stevens, B., M.~Satoh, L.~Auger, J.~Biercamp, C.~Bretherton, X.~Chen,
  P.~Dueben, F.~Judt, M.~Khairoutdinov, D.~Klocke, C.~Kodama, L.~Kornblueh,
  S.-J. Lin, W.~Putman, R.~Shibuya, P.~Neumann, N.~Roeber, B.~Vanniere, P.-L.
  Vidale, N.~Wedi, and L.~Zhou (Submitted), {DYAMOND}: The {DYnamics} of the
  {A}tmospheric general circulation {M}odeled {O}n {N}on-hydrostatic
  {D}omains., \textit{Prog. Earth Planet. Sci.}

\bibitem[{\textit{Stratton and Stirling}(2012)}]{Stratton2012-iu}
Stratton, R.~A., and A.~J. Stirling (2012), Improving the diurnal cycle of
  convection in {GCMs}, \textit{Q.J.R. Meteorol. Soc.}, \textit{138}(666),
  1121--1134, \doi{10.1002/qj.991}.

\bibitem[{\textit{Woelfle et~al.}(2018)\textit{Woelfle, Yu, Bretherton, and
  Pritchard}}]{Woelfle2018-gi}
Woelfle, M.~D., S.~Yu, C.~S. Bretherton, and M.~S. Pritchard (2018),
  Sensitivity of coupled tropical pacific model biases to convective
  parameterization in {CESM1}, \textit{J. Adv. Model. Earth Syst.},
  \textit{10}(1), 126--144, \doi{10.1002/2017MS001176}.

\bibitem[{\textit{Yanai et~al.}(1973)\textit{Yanai, Esbensen, and
  Chu}}]{Yanai1973-gt}
Yanai, M., S.~Esbensen, and J.-H. Chu (1973), Determination of bulk properties
  of tropical cloud clusters from {Large-Scale} heat and moisture budgets,
  \textit{J. Atmos. Sci.}, \textit{30}(4), 611--627,
  \doi{10.1175/1520-0469(1973)030<0611:DOBPOT>2.0.CO;2}.

\bibitem[{\textit{Zhang and Wang}(2006)}]{Zhang2006-iy}
Zhang, G.~J., and H.~Wang (2006), Toward mitigating the double {ITCZ} problem
  in {NCAR} {CCSM3}, \textit{Geophys. Res. Lett.}, \textit{33}(6), 4641,
  \doi{10.1029/2005GL025229}.

\end{thebibliography}






\end{document}